\documentclass[12pt]{article}

\usepackage{setspace}
\usepackage{amsmath}
\usepackage{amsfonts}
\usepackage{amssymb}
\usepackage{amsthm}
\usepackage{xcolor}
\usepackage{wrapfig}
\usepackage{srcltx}
\usepackage{epsfig}
\usepackage{subcaption}
\usepackage{makecell}
\usepackage{hyperref}
\def\bE{{\mathbf{E}}}
\def\Prob{{\mathrm{Prob}}}
\def\VI{{\mathrm{VI}}}
\def\bR{{\mathbf{R}}}
\def\cX{{\cal X}}

\def\cR{{\cal R}}

\def\Dom{{\mathrm{Dom}\,}}
\def\cB{{\cal B}}
\def\mfB{{\mathfrak{B}}}
\def\mfZ{{\mathfrak{Z}}}

\def\Col{\mathrm{Col}}
\def\Row{\mathrm{Row}}

\def\myE{\widetilde{E}}

\newcommand{\be}{\begin{eqnarray}}
\newcommand{\ee}[1]{\label{#1}\end{eqnarray}}

\newtheorem{proposition}{Proposition}[section]
\newtheorem{definition}{Definition}[section]

\newcommand{\yx}[2]{{}{\color{red}#2}} 

\def\cH{{{\cal H}}}
\def\cZ{{{\cal Z}}}

\newcommand{\hide}[1]{}

\title{``Generalized'' generalized linear models: \\Convex estimation and online bounds}

\author{Anatoli Juditsky\thanks{LJK, Universit\'e Grenoble Alpes, B.P. 53, 38041 Grenoble Cedex 9, France,	
{\tt anatoli.juditsky@imag.fr}}
\and Arkadi Nemirovski\thanks{Georgia Institute
 of Technology, Atlanta, Georgia
30332, USA, {\tt nemirovs@isye.gatech.edu}}\and
Yao Xie\thanks{Georgia Institute
 of Technology, Atlanta, Georgia
30332, USA, {\tt yao.xie@isye.gatech.edu}}
\and
Chen Xu\thanks{Georgia Institute
 of Technology, Atlanta, Georgia
30332, USA, {\tt cx9711@gatech.edu}\newline
The work of Xu and Xie is partially supported by an NSF CAREER CCF-1650913, and NSF DMS-2134037, CMMI-2015787, CMMI-2112533, DMS-1938106, and DMS-1830210.} 
}

\begin{document}

\maketitle

\begin{abstract}

We introduce a new computational framework for estimating parameters in generalized generalized linear models (GGLM), a class of models that extends the popular generalized linear models (GLM) to account for dependencies among observations in spatio-temporal data. The proposed approach uses a monotone operator-based variational inequality method to overcome non-convexity in parameter estimation and provide guarantees for parameter recovery. The results can be applied to GLM and GGLM, focusing on spatio-temporal models. We also present online instance-based bounds using martingale concentrations inequalities. Finally, we demonstrate the performance of the algorithm using numerical simulations and a real data example for wildfire incidents.

\end{abstract}

\section{Introduction}

We aim to perform parameter estimation for generalized generalized linear models (GGLM) and provide methods for quantifying the uncertainty in the estimated parameters. GGLM can be thought of as an extension of the widely used generalized linear models (GLM) \cite{nelder1972generalized,mccullagh2018generalized,agresti2003categorical}, which offer a versatile set of models for statistical analysis of data when we have a set of measurements or counts accompanied by relevant contextual information. GLM encompasses a broad range of models that map input/response variables to predictors, such as logit and probit models for quantal response data and log-linear models for count response data. However, standard GLM models assume that the observations are independent, which precludes modeling of "data exhibiting the autocorrelations of time series and spatial processes," as noted in \cite{mccullagh2018generalized}. GGLM extends GLM to account for dependencies among observations in spatio-temporal data.

This paper presents a computational framework for the ``generalized'' generalized linear model (GGLM) using a new convex formulation by solving the monotone operator-based variational inequalities (VI) \cite{juditsky2019signal}.  In particular, we emphasize the use of GGLM for spatio-temporal models, although the results can still be applied to GLM if we remove the spatial and temporal dependence. We aim to build the bridge between the generalized general linear model and the variational inequality in optimization. This leads to efficient computational procedures that overcome the non-convexity in directly fitting the problem via least-square formulations. In some cases, the new VI approach will also enable performance guarantees for maximum likelihood estimates (MLE) for model parameters.

The conventional approach for estimating parameters in GLM models involves using weighted least-squares (WLS) \cite{nelder1972generalized,mccullagh2018generalized,agresti2003categorical}. Although WLS is commonly used in practice and offers a convenient estimation strategy, it does not provide precise guarantees for parameter recovery. In contrast, the current study proposes a VI monotone operator recovery method that offers guaranteed parameter recovery and online concentration bounds for GGLM. Notably, these results are applicable to the classic GLM and its spatio-temporal extension.

Moreover, our approach avoids the non-convexity caused by non-linear link functions. In order to construct a convex VI problem, the structural assumptions to be imposed on the model are essentially weaker than those resulting in convex ML-based or LS-based problems. The VI problem we solve may or may not be equivalent to a convex minimization problem. For instance, we will show that the VI problem we construct will reduce to the LS problem under linear link functions. Moreover, the proposed GGLM emphasizes spatio-temporal dependencies. Thus, the observations are no longer i.i.d. and can have complex dependencies. Therefore, martingale concentrations are developed to quantify the recovery error.

The rest of the paper is organized as follows. Section \ref{GGLM} introduces the GGLM model, recovery algorithm, and performance metrics. Section \ref{sec:guarantee} presents performance guarantees for recoveries based on solving monotone operator-based variational inequalities. Section \ref{sec:app} discusses how to use the framework in a special case: spatio-temporal processes, followed by two representative examples for Bernoulli processes (Section \ref{sec:ber}) and Poisson processes (Section \ref{sec:Poisson}). Finally, we conclude the paper with discussions.

\vspace{.1in}
\par\noindent
{\bf Notation:} The notations of the paper is standard. All linear spaces we work in are $\ bR^n$, with different dimensions $n$, equipped with the standard Euclidean structure. Sometimes the vectors from $\bR^n$ in question have additional structure -- they are represented as one-, two-, or even three-dimensional arrays of blocks of common size. In all cases, we assume that the entries in a vector from $\bR^n$ under consideration independently of the additional block structure, if any, are assigned serial numbers. This allows us to write inner products as $x^Ty$, to identify a linear mapping from $\bR^n$ into $\bR^m$ with $m\times n$ matrix, its conjugate -- with the transpose of this matrix, etc. Let $[x]_i$ denote the $i$th entry of a vector $x$, and $[X]_{ij}$ denote the $(i, j)$th entry of a matrix $X$; $I_m$ denotes an identity matrix of size $m\times m$.

\section{``Generalized'' Generalized Linear Models}
\label{GGLM}

We start by presenting the main modeling framework, which we will demonstrate using examples of its wide applicability. In statistics, the generalized linear model (GLM) \cite{mccullagh2018generalized} is a flexible generalization of the Gaussian model: where the response variable is related to the predictor by a linear function contaminated by additive Gaussian noise. The GLM generalizes linear regression by allowing the response variable to be related to the predictors via a non-linear link function, which determines the mean and variance (dispersion) of the response.

\subsection{Situation and goal}\label{GGLM_setup}

%
%
%
%

Consider the  ``Generalized'' Generalized Linear Models (GGLM) as follows
\begin{quote}
At time $t=1,2,...$ we observe random pair $\upsilon_t=(\zeta_t,H_t)$, where random vector $\zeta_t$ is the response that takes values in a known subset $\cZ_t$ of some $\mfZ_t=\bR^{n_t}$, $H_t$ is random $\kappa\times n_t$ matrix that contains the predictor variables taking values in a known subset $\cH_t$ of the space $\bR^{\kappa\times n_t}$ of $\kappa\times n_t$ matrices. Similar to the GLM model, we specify the (conditional) mean of the response variable. For every $t=1,2,...$ the conditional, $\upsilon^{t-1}=(\upsilon_1,...,\upsilon_{t-1})$ (assuming $\upsilon^0 = 0$) {\sl and} $H_t$ given, distribution of $\zeta_t$ has expectation
\begin{equation}
\bE_{|\upsilon^{t-1},H_t}\{\zeta_t\}=\Phi_t(H_t^T\beta),
\label{cond_prob}
\end{equation}
where
\begin{itemize}
\item $\Phi_t(\cdot):\Dom\Phi_t\to \mfZ_t$ --- the $t$-th known link function --- is a continuous monotone mapping defined on a closed convex domain $\Dom\Phi_t\subset \mfZ_t$, monotonicity meaning that
    $$
    [\Phi_t(x)-\Phi_t(y)]^T[x-y]\geq0\,\,\forall x,y\in\Dom\Phi_t;
    $$
\item $\beta$ is the unknown vector of parameters taking values in a convex compact subset $\cB$ of $\mfB = \bR^\kappa$; $\kappa$ is the number of parameters.
\item $\cH_t$ and $\mfB$ are such that $H_{t}^T x\in\Dom\Phi_t$ for all $H_{t}\in\cH_{t}$, $x\in\mfB$, and $t$.
\end{itemize}
We call the situation {\sl stationary}, if $n_t$, $\mfZ_t$, $\cZ_t$, $\cH_t$ and $\Phi_t$  are independent of $t$. Our goal is to recover $\beta$ from observations $\upsilon^N=(\upsilon_1,...,\upsilon_N)$, where $N$ is a given time horizon. A special case is when the mapping (or link function) $\Phi(\cdot)$ is the same for each entry of $\zeta_t$, for instance, the sigmoid function $\phi(z) = 1/(1+e^{-z})$, such that $\Phi_t(\zeta) = [\phi([\zeta]_1), \ldots, \phi([\zeta]_n)]$ for $\zeta \in \bR^n$ and $[\zeta]_i$ is the $i$th entry (see \cite{mccullagh2018generalized} for examples of link functions). Later on, we will also further assume that the entries of $\zeta_t$ are conditionally independent, given $\upsilon_t$ and $H_t$.
\end{quote}

\begin{figure}
\begin{center}
\includegraphics[width = 0.6\textwidth]{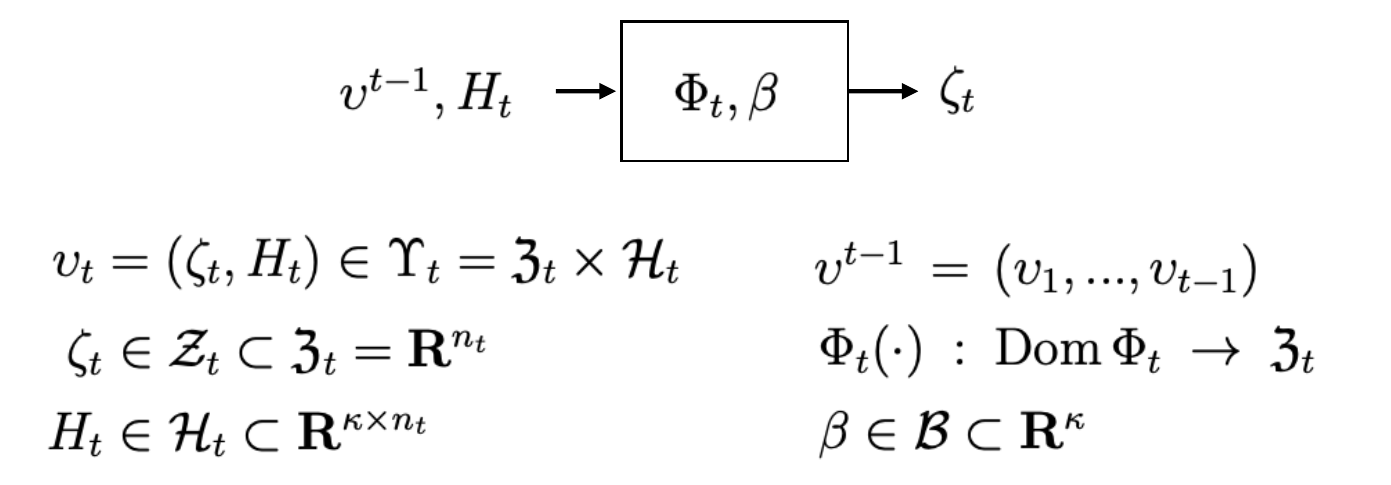}
\vspace{.1in}
\caption{Illustration of the Generalized Generalized Linear Model (GGLM), where $\zeta_t$ and $H_t$ are a sequence of predictors and response variables, respectively. The link function $\Phi_t$ is known and the parameters for predictors $\beta$ are unknown.}

\end{center}
\end{figure}

\subsection{Connection with GLM}\label{sec:GLM}

The stationary version of this model resembles the {\it generalized linear model (GLM)} \cite{mccullagh2018generalized}, when the observations $\upsilon_t=(H_t,\zeta_t)$, $t = 1, 2, \ldots $ are i.i.d., $H_{t} \in \bR^\kappa$ corresponds to the predictors (regressors), and $\zeta_t \in \bR$ corresponds to the response variable (in GLM it is typically a scalar.) 
GLM specifies the model using a {\it systematic component} that $\bE[\zeta] = \phi(H_t^T \beta)$  for a monotonic differentiable link function $\phi$ (induced by the distribution of $\zeta$), and a {\it random component} assuming the response follows a distribution that usually comes from an exponential family. Thus, observations are realizations of a random variable $\zeta$, whose distribution $\psi$ takes the form of
\begin{equation}
\zeta\sim \psi(\zeta; \theta, \varrho) = \exp\{(\zeta \theta - B(\theta))/A(\varrho) + D(\zeta, \varrho)\}, \label{exponential_family}
\end{equation}
for some specific function $A(\cdot)$, $B(\cdot)$, and $D(\cdot)$; the natural parameter is $\theta$ and the nuisance parameter is $\varrho$. If $\varrho$ is known, this is a single-parameter exponential-family model.  
%

It can be shown that the mean $\bE[\zeta] = B'(\theta)$ and variance $\mbox{Var}(\zeta) = B''(\theta).$ Except for the Gaussian distribution, the variance is related to the mean for all other exponential family distributions. Thus, one can define GLM induced by an exponential family distribution by specifying the conditional mean in the form of \eqref{cond_prob}.  GLM typically assumes that $\theta$ linearly depends on the predictors (regressors), and the link function can be chosen according to the assumed distribution. Some examples are as follows: (i) Gaussian distribution $\zeta\sim \mathcal N(\mu, \sigma^2)$,  $\theta = \mu$,  $\varrho = \sigma^2$, $A(\varrho) = \varrho$, $B(\theta) = \theta^2/2$, $D(\zeta, \varrho) = -\frac 12 \{\zeta^2/\sigma^2 + \log (2\pi\sigma^2)\}$. The mean $\bE [\zeta] = B'(\theta) = \theta = \mu$ and the link function is an identity (GLM reduces to the linear model in this case). (ii) Bernoulli distribution $\zeta \in \{0, 1\}$ with parameter $p \in [0, 1]$. The density function is given by
\[\psi(\zeta; p) = \zeta^p(1-\zeta)^{1-p} = \exp\{\zeta \theta - \log (1+e^\theta)\},
\]
where the natural parameter $\theta = \log \frac{p}{1-p}$, $\varrho = 1$, $A(\varrho)=1$, $B(\theta)=\log(1+e^\theta)$; $\bE[\zeta] = B'(\theta) = 1/(1+e^{-\theta}) = p$ and $\mbox{Var}[\zeta] = B''(\theta) = e^{\theta}/(1+e^{\theta})^2 = p(1-p)$; the link function $\phi(z)$ is a sigmoid function. (iii) Poisson distribution $\zeta = 0, 1, 2, \ldots$ with parameter $\lambda$. The density function is given by
\[\psi(\zeta; \lambda) = \lambda^\zeta e^{-\lambda}/\zeta! = \exp\{ \zeta \log \lambda -\lambda  -\log (\zeta!)\}\]
with natural parameter $\theta = \log \lambda$, $B(\theta) = e^\theta$. Thus, $\bE[\zeta] = B'(\theta) = e^\theta = \lambda$ and $\mbox{Var}[\zeta] = B''(\theta) = e^{\theta} = \lambda$, and the link function is the log function.


\section{Convex parameter recovery}
\label{sec:paramrec}

 The problem we address here is recovering the parameters $\beta$ of the GGLM model from observations. To this end, we intend to use a {\sl Variational Inequality} (VI) oriented procedure; when the link functions are monotone (as it was assumed when presenting a GGLM model), the resulting procedure is computationally efficient. In contrast, the computation is not always efficient for more traditional recovery procedures using classical loss functions or based on Maximum Likelihood, which requires the concavity of the log-likelihood function.

\vspace{.1in}
\noindent
{\bf Monotone vector fields and variational inequalities.} We start by introducing the necessary background information.

\begin{definition}[Monotone vector field.]
A vector field $F: \cX\rightarrow   \bR^N$ defined on a nonempty convex subset $\cX$ of $\bR^N$ is called monotone, if $\langle F(x) - F(y), x- y \rangle \geq 0$ whenever $x, y\in \cX$.
\end{definition}
The set $\cX$ and a monotone on $\cX$ vector field $F$ specify {\sl Variational Inequality
$$
\hbox{Find $z\in\cX:$}\ \langle F(w), w-z\rangle\geq 0, \quad \forall w\in\cX.\eqno{\VI[F, \cX]}
$$
Vectors $z$ satisfying the requirement just defined are called weak solutions to $\VI[F, \cX]$.}
A vector $z\in \cX$ is called a {\it strong solution} to $\VI[F, \cX]$ when
\[\langle F(z), w-z\rangle\geq0, \quad \forall w\in \cX.\]
From the monotonicity of $F$ it follows that every strong solution is a weak one.
When the monotone vector field $F$ is continuous, the weak solutions to $\VI[F,\cX]$ are the same as strong solutions. Weak solutions definitely exist whenever $\cX$ is a nonempty convex compact set.
\par
 Some basic examples of monotone vector fields are\\
 ---  the gradient field $\nabla f(x)$ of continuously differentiable convex function $f:\cX\to\bR$;
\\
--- the vector field $F(u,v)=[\nabla_uf(u,v);-\nabla_vf(u,v)]$ of continuously differentiable convex in $u$ and concave in $v$ function $f(u,v):\cX:=U\times V\to\bR$.
\par\noindent
In these cases, the weak$\equiv$strong solutions to $\VI[F,\cX]$ are, respectively, the minimizers of $f$ on $\cX$ and the saddle points ($\min$ in $u$, $\max$ in $v$) of $f$ on $U\times V$.

\vspace{.1in}
\par\noindent
{\bf Application to GGLM case.} Given a GGLM from Section \ref{GGLM},  we associate with an observation
\[\upsilon^N=\{\upsilon_\tau=(\zeta_\tau,H_\tau),1\leq\tau\leq N\}\in \overbrace{[\cZ_1\times\cH_1]}^{\Upsilon_1}\times \ldots \times \overbrace{[\cZ_N\times\cH_N]}^{\Upsilon_N},\] a collection of scalar weights $\lambda_t > 0$, $1\leq t\leq N,$ which should be deterministic functions of $(\upsilon^{t-1},H_t)$: $\lambda_t=\Lambda_t(\upsilon^{t-1},H_t)$ and two vector fields. First, the observable field
\begin{equation}
F_{\upsilon^N}(z)={1\over N}\sum_{t=1}^N\Lambda_t^{-1}(\upsilon^{t-1},H_t) H_{t}\left[\Phi_t\left(H_{t}^Tz\right)-\zeta_t\right]:\cB\to\mfB,
\label{observed_F}
\end{equation}
and second, the unobservable field which is a ``population'' version of $F_{\upsilon^N}(\cdot)$,
\begin{equation}
\overline{F}_{\upsilon^N}(z)={1\over N}\sum_{t=1}^N\Lambda_t^{-1}(\upsilon^{t-1},H_t)H_{t}\left[\Phi_t\left(H_{t}^Tz\right)-\Phi_t\left(H_{t}^T\beta\right)\right]:\cB\to\mfB.
\label{population_F}
\end{equation}
Note that in our situation, both of these vector fields are well-defined and continuous on $\cB$; they clearly are monotone along with $\Phi_t$.\footnote{We have used the elementary observation that if $x\mapsto Ax+b$ is affine mapping which maps a convex domain $X\subset\bR^N$ into the domain $\Dom G\subset\bR^M$ of a monotone vector field $G$, then the vector field $A^TG(Ax+b)$ is monotone on $X$.}
 In addition, $\cB$ is compact.

 Consequently, the VI is defined using the observed field \eqref{observed_F}:
$$
\hbox{Find\ }z\in\cB: \langle F_{\upsilon^N}(x),x-z\rangle \geq 0,\,\,\forall x\in\cB \eqno{\VI(F_{\upsilon^N},\cB)}.
$$
has solutions, and these weak solutions are the same as the strong solutions to the VI.
Our estimate of $\beta$, denoted by $\widehat \beta$, will be the weak$\equiv$strong solution to the variational inequality $\VI(F_{\upsilon^N},\cB)$.\\

\noindent
{\bf Special case.}
When the monotone vector mappings $\Phi_t:\Dom\Phi_t\to\mfZ_t$ are the gradient fields of  convex functions $\mathfrak{F}_t(\cdot):\Dom\Phi_t\to\bR$, the vector field $F_{\upsilon^N}$ is the gradient field of the convex function
\begin{equation}\label{eq4}
\mathfrak{F}_{\upsilon^N}(x)={1\over N}\sum_{t=1}^N\Lambda^{-1}_t(\upsilon^{t-1},H_t)\left[\mathfrak{F}_t(H_{t}^Tx)-x^TH_{t}\zeta_t\right]
\end{equation}
so that the
 solutions to $\VI(F_{\upsilon^N},\cB)$ are exactly the optimal solutions to the convex minimization problem
\begin{equation}\label{eq5}
\min_{x\in\cB}\mathfrak{F}_{\upsilon^N}(x).
\end{equation}
In particular, in the case of the identity link $\Phi_t(\eta_{t})\equiv\eta_{t}$, $\eta_{t}\in\mfZ_{t}$, we have $\Phi_t(\eta_t)=\nabla \left[{1\over 2}\eta_t^T\eta_t\right]$, thus $\mathfrak{F}_t(x) \propto \frac 1 2 x^T x$, $\Lambda^{-1}_t(\upsilon^{t-1}, H_t)=1$, and it can be verified that the proposed recovery becomes just the Least Squares:
\begin{equation}\label{eq_eg}
\min_{x\in\cB} {1\over {2N}}\sum_{t=1}^N \left\| H_{t}^Tx-\zeta_t\right\|^2.
\end{equation}

\par\noindent
{\bf Maximum Likelihood estimation.}
Assuming that the density, taken w.r.t. some reference measure, of the conditional, $\upsilon^{t-1}$ and $H_t$ given, distribution of $\zeta_t$ is a known function of the form $\psi_t(\zeta_t; H_{t}^T \beta)$ as specified in Section \ref{sec:GLM}, we could try to recover $\beta$ by maximizing the likelihood of what we have observed, that is, by solving the optimization problem
$$
\max_{x\in\cB}{1\over N}\sum_{t=1}^N\ln(\psi_t(\zeta_t; H_{t}^T x)).
$$
In order for this approach to be computation-friendly, the problem should be convex. When the link function is derived from the exponential family, the maximum likelihood problems are usually convex. In such cases, we can use the gradient of the log-likelihood function as the monotone operator in VI, and it can be shown that the weights $\Lambda_t^{-1}(\upsilon^{t-1}, H_t)$ in \eqref{observed_F} can be related to terms in the gradient of the log-likelihood function.

\vspace{.1in}
\noindent
{\bf Remark.} For GLM, the maximum likelihood parameter estimation is typically done by Iterative Weighted Least Squares (IWLS)  \cite{mccullagh2018generalized}. One may wonder whether it relates to the proposed VI-based approach using \eqref{observed_F}. However, they are not the same. IWLS is a heuristic iterative procedure motivated by Fisher's scoring method (a variant of Newton's step using the expected value of the Hessian matrix). IRLS iteratively fits the adjusted response variable with respect to predictors using the weights that depend on the fitted parameter values from the previous iterations. The parameter recovery guarantee for IWLS is unclear.
\hfill
$\Box$


\section{Performance Guarantees}\label{sec:guarantee}

We now establish results underlying theoretical guarantees for parameter recovery. We focus on the results for parameters recovered using VI.

\subsection{Concentration of empirical vector field}
\label{sec:concentration}

An important observation is as follows:
\begin{center}
$\beta$ is the root of $\overline{F}_{\upsilon^N}(\cdot)$ in $\cB$: $\overline{F}_{\upsilon^N}(\beta) = 0$,
\end{center}
and
\[
F_{\upsilon^N}(\beta)=F_{\upsilon^N}(\beta)-\overline{F}_{\upsilon^N}(\beta)={1\over N}\sum_{t=1}^N\Lambda_t^{-1}(\upsilon^{t-1},H_t)H_{t}\left[\Phi_t\left(H_{t}^T\beta\right)-\zeta_t\right].
\]
What is under the summation is {\it martingale-difference}, since
\[
\bE_{|\upsilon^{t-1},H_t}[F_{\upsilon^N}(\beta)-\overline{F}_{\upsilon^N}(\beta)]
= {1\over N}\Lambda_t^{-1}(\upsilon^{t-1},H_t)H_{t}\cdot\bE_{|\upsilon^{t-1},H_t}\left[\Phi_t\left(H_{t}^T\beta\right)-\zeta_t\right] = 0;
\]
we have used (\ref{cond_prob}). We first derive the following concentration inequality, which is useful in establishing our results for parameter recovery and online bounds for GGLM. We will illustrate their usage for spatial-temporal processes in Section \ref{sec:app}.
\begin{proposition}[Concentration of empirical vector field]\label{cor1}
Assume that for every $t$, every $\upsilon^{t-1}\in \Upsilon_1\times\ldots\times \Upsilon_{t-1}$ and every $H_t\in\cH_t$ the conditional, $\upsilon^{t-1},H_t$ given, distribution $P_{|\upsilon^{t-1},H_t}$ of $\zeta_{t}$ satisfies
\begin{equation}\label{eq2}
\ln\left(\bE_{\zeta_{t}\sim P_{|\upsilon^{t-1},H_t}}\left\{\exp\{h^T\zeta_{t}\}\right\}\right)\leq h^T\Phi_t\left(H_{t}^T\beta\right)+\Psi_{t,\upsilon^{t-1},H_t}(\|P_th\|_1),\,\,\forall h\in\mfZ_t
\end{equation}
where the rate functions $\Psi_{t,\upsilon^{t-1},H_t}(r): \bR_+\to\bR$ are continuous nondecreasing functions, and $P_t$ are known deterministic matrices.
\par
Let $E_{t}$, $t=1,2,...,$ be $m\times n_t$ matrices which are deterministic functions of $(\upsilon^{t-1},H_t)$: $E_{t}=E_{t}(\upsilon^{t-1},H_t)$, and let
$$
G_{\upsilon^N}={1\over N}\sum_{t=1}^NE_{t}[\Phi_t(H_{t}^T\beta)-\zeta_t].
$$
Then
\par
{\rm (i)} For $\upsilon_\tau\in\Upsilon_\tau$, $1\leq\tau\leq N$, $1\leq j \leq m$, let $\Theta_{sj}(\upsilon^{s-1},H_s)$ be the $\|\cdot\|_1$-norm of $j$-th column in $P_sE^T_{s}$, $s = 1, 2, \ldots, N$. Given a finite set $\Gamma=\{\alpha_i>0,1\leq i\leq K\}$ on the positive ray, and a tolerance $\epsilon\in(0,1)$, for every $j\leq m$ the  probability of the event
\begin{equation}\label{eqevent}
\bigg\{\upsilon^N:|[G_{\upsilon^N}]_j|> \underbrace{\min\limits_{1\leq i\leq K}\bigg[\alpha_i\ln(2 K/\epsilon)+\alpha_i\sum_{t=1}^N\Psi_{t,\upsilon^{t-1},H_t}(\alpha_i^{-1}N^{-1}\Theta_{tj}(\upsilon^{t-1},H_t))}_{\delta_j(\upsilon^N)}\bigg]\bigg\}
\end{equation}
is at most $\epsilon$.

\par
{\rm (ii)} Assume that $\Psi_{t,\upsilon^{t-1},H_t}(\cdot)\leq \Psi(\cdot)$ for all $t$, where $\Psi(\cdot):\bR_+\to\bR_+$ is a continuous nondecreasing ``worst case'' rate function, and that for some $\Theta<\infty$,
$$\max_{i} \|\Col_i[P_{t}E_{t}^T]\|_1 \leq \Theta, \mbox{ whenever }  \upsilon_s\in \Upsilon_s, 1\leq s\leq t, \forall t.$$
(from now on, $\Col_j[A]$ stand for $j$-to column, and $\Row_i[A]$ -- for the transpose of $i$-th row in matrix $A$).
Then for every $\delta\geq0$ one has
\begin{equation}\label{thenprob}
\Prob\{\|G_{\upsilon^N}\|_\infty>\delta\}  \leq 2m\exp\{-N\Psi_{*}(\delta/\Theta)\},
\end{equation}
where for $s\geq0$
\begin{equation}
\Psi_{*}(s) = \sup_{\gamma\geq0}  \left[\gamma s-\Psi(\gamma)\right]\geq0,
\end{equation}
is the Fenchel-Legendre transformation of  $\Psi(\cdot)$.
\end{proposition}

\noindent
{\bf Remark.} Assume that we are in the situation of the item (ii) and are given $\varepsilon\in(0,1)$, and let us look at the upper bound of the $(1-\varepsilon)$-quantile of $\|G_{\upsilon^N}\|_\infty$  we could get
from (\ref{thenprob}).
In order for (\ref{thenprob}) to imply that $\Prob\{\|G_{\upsilon^N}\|_\infty>\delta\}\leq\varepsilon$ we need to ensure that
$$
\sup_{\gamma>0}\left[\gamma [\delta/\Theta]-\Psi(\gamma)\right]\geq N^{-1}\ln(2m/\varepsilon),
$$
or, which is the same (set $\gamma=\Theta/(N\alpha)$), that
$$
\sup_{\alpha>0}\left[\delta-\alpha N\Psi(\Theta/(N\alpha))-\alpha\ln(2m/\varepsilon)\right]\geq0.
$$ The smallest $\delta=\widehat{\delta}$ satisfying this requirement is
$$
\widehat{\delta}=\inf_{\alpha>0}\left[\alpha \ln\left(2m/\varepsilon\right)+\alpha N\Psi(N^{-1}\Theta/\alpha)\right].
$$


On the other hand, we are in the situation when $\Theta_{tj}(\upsilon^{t-1},H_t)\leq\Theta$ for all $t$, $j$ and all $\upsilon^{t-1}\in\Upsilon^{t-1}$, $H_t\in\cH_t$. It follows that given $\delta>\widehat{\delta}$, selecting $\bar{\alpha}>0$ in such a way that
$$\left[\bar{\alpha} \ln\left(2m/\varepsilon\right)+\bar{\alpha} N\Psi(N^{-1}\Theta/\bar{\alpha})\right]\leq\delta$$ and specifying $\Gamma=\{\bar{\alpha}\}$, $\epsilon=\varepsilon/m$, the right-hand side in the inequality in (\ref{eqevent}) would, for every $j\leq m$ be $\leq\delta$, implying, via the union bound, that the probability of the event
$$
\|G_{\upsilon^N}\|_\infty>\delta
$$
is $\leq\varepsilon$.
 Thus, in the case of (ii), the result of (i) with properly selected singleton $\Gamma$ would result in the upper bound on the $(1-\varepsilon)$-quantile of $\|G_{\upsilon^N}\|_\infty$
which can be made arbitrarily close to the bound yielded by (ii). \qed

\vspace{.1in}
The rationale beyond (i) is that it results in {\sl online bounds} $\delta_j(\upsilon^N)$ on the $(1-\epsilon)$-quantile of $\|G_{\upsilon^N}\|_\infty$, meaning that {\sl when $\Psi_{t,\upsilon^{t-1},H_t}$ is observable}, the bound is an observable function of $\upsilon^N$ such that the probability for $|[G_{\upsilon^N}]_j|$ to exceed this bound is at most $\epsilon$. This online bound adjusts itself to
the observed rate functions $\Psi_{t,\upsilon^{t-1},H_t}(\cdot)$ and observed values of $\Theta_{tj}(\upsilon^{t-1},H_t)$, in contrast to the deterministic bound $\widehat{\delta}$ oriented on the worst case rate function and a priori upper bound $\Theta$ on $\Theta_{tj}(\upsilon^{t-1},H_t)$'s. Note that the price for cardinality $K$ of $\Gamma$ (the larger $K$, the stronger ``adaptive abilities'' of the online bound $\delta_j(\cdot)$) is low -- $K$ is under the log.


\par\noindent
\\
{\sl Proof} of Proposition \ref{cor1}.
Let us set
\[
\begin{array}{rcl}
S_t(\upsilon^t)&=&\sum_{s=1}^t{1\over N}E_{s}\left[\Phi_s\left(H_{s}^T\beta\right)-\zeta_s\right],\\
R_t(\upsilon^t;g)&=&\sum_{s=1}^t\Psi_{s,\upsilon^{s-1},H_s}(N^{-1}\|P_sE_{s}^Tg\|_1)\quad[g\in\bR^{m}],\\
\end{array}
\]
and let $\bE_t$ stand for  expectation over $\upsilon^t$, and $\bE^+_t$ stand for expectation over $(\upsilon^t,H_{t+1})$.
\par(i):  Let us fix $j$, and let $g=\chi e_j$, where $e_j$ is $j$-th standard basic orth in $\bR^m$ and $\chi=\pm1$, and $\gamma\geq0$, and let
\begin{equation}\label{eq446}
\begin{array}{rcl}
\Sigma_t(\upsilon^t)&=&\sum_{s=1}^t\bigg[{1\over N}\gamma g^TE_{s}\left[\Phi_s\left(H_{s}^T\beta\right)-\zeta_s\right]
-\Psi_{s,\upsilon^{s-1},H_s}(\|\gamma N^{-1}P_sE_{s}^Tg\|_1)\bigg]\\
&=& \gamma g^TS_t(\upsilon^t)-R_t(\upsilon^t;\gamma g).
\end{array}
\end{equation}
We have
\[
\begin{array}{l}
\bE_{t+1}\left\{\exp\{\Sigma_{t+1}(\upsilon^{t+1})\}\right\}\\
= \bE_t^+\bigg\{\bE_{\zeta_{t+1}\sim P_{|\upsilon^{t},H_{t+1}}}\bigg\{\exp\bigg\{\Sigma_t(\upsilon^t)-\Psi_{t+1,\upsilon^t,H_{t+1}}(\|\gamma N^{-1}P_{t+1}E_{t+1}^Tg\|_1)\\
\multicolumn{1}{r}{+\gamma N^{-1}[E_{t+1}^Tg]^T[{\Phi}_{t+1}\left(H_{t+1}^T\beta\right)-\zeta_{t+1}]\bigg\}\bigg\}\bigg\}}\\
=\bE_t^+\bigg\{\exp\bigg\{\Sigma_t(\upsilon^t)-\Psi_{t+1,\upsilon^t,H_{t+1}}(\|\gamma N^{-1}P_{t+1}E_{t+1}^Tg\|_1)\bigg\}\\
\multicolumn{1}{r}{\times\underbrace{\bE_{\zeta_{t+1}\sim P_{|\upsilon^{t},H_{t+1}}}\bigg\{\exp\big\{
\gamma N^{-1}[E_{t+1}^Tg]^T[{\Phi}_{t+1}\left(H_{t+1}^T\beta\right)-\zeta_{t+1}]\big\}\bigg\}}_{\leq
\exp\{\Psi_{t+1,\upsilon^t,H_{t+1}}(\|\gamma N^{-1}P_{t+1}E_{t+1}^Tg\|_1)\}\hbox{\tiny\ by (\ref{eq2})}}\bigg\}}
\\
\leq \bE_t^+\left\{\exp\{\Sigma_t(\upsilon^t)\}\right\}=\bE_t\left\{\exp\{\Sigma_t(\upsilon^t)\}\right\}.\\
\end{array}
\]
This reasoning works for all $t\geq0$, with the empty sum $\Sigma_0(\upsilon^0)$, as usual, identically equal to 0. Consequently,
$
\bE_{\upsilon^N}\left\{\exp\{\Sigma_N(\upsilon^N)\}\right\}\leq 1,
$
or, which is the same (note that $G_{\upsilon^N}=S_N(\upsilon^N)$):
\[
\bE_{\upsilon^N}\left\{\exp\{\gamma g^TG_{\upsilon^N}-R_N(\upsilon^N;\gamma g)\}\right\}\leq 1,
\]
whence for every $\delta\geq0$ one has by Markov inequality that
$$
\Prob\left\{\gamma g^TG_{\upsilon^N}-R_N(\upsilon^N;\gamma g)>\gamma \delta\right\}\leq  \exp\{-\gamma\delta\}.
$$
Setting $\gamma=1/\alpha$ and recalling what $R_N$ is, we get
$$
\Prob\left\{g^TG_{\upsilon^N}>\alpha\sum_{t=1}^N\Psi_{t,\upsilon^{t-1},H_t}(
\|\alpha^{-1}N^{-1}P_tE_{t-1}^Tg\|_1)+\delta\right\}\leq \exp\{-\delta/\alpha\},
$$
whence, due to monotonicity of $\Psi_{t,\upsilon^{t-1},H_t}(\cdot)$ and the fact that $\|P_tE_{t}^Tg\|_1\leq\Theta_{tj}(\upsilon^{t-1},H_t)$, we have
$$
\Prob\left\{g^TG_{\upsilon^N}>\alpha\sum_{t=1}^N\Psi_{t,\upsilon^{t-1},H_t}(\alpha^{-1}N^{-1}\Theta_{tj}(\upsilon^{t-1},H_t))+\delta\right\}\leq \exp\{-\delta/\alpha\}.
$$
Using the union bound, we get
\begin{equation}\label{eq333}
\Prob\left\{|[G_{\upsilon^N}]_j|>\delta+\alpha\sum_{t=1}^N\Psi_{t,\upsilon^{t-1},H_t}(\alpha^{-1}N^{-1}\Theta_{tj}(\upsilon^{t-1},H_t))\right\}\leq 2\exp\{-\delta/\alpha\}.
\end{equation}
For $1\leq i\leq K$, let us set $\delta_i=\alpha_i\ln(2K/\epsilon)$, so that by (\ref{eq333}) the probability of the event
$$
{\cal E}_i=\{\upsilon^N: |[G_{\upsilon^N}]_j|>\alpha_i\ln(2 K/\epsilon)+\alpha_i\sum_{t=1}^N\Psi_{t,\upsilon^{t-1},H_t}(\alpha_i^{-1}N^{-1}\Theta_{tj}(\upsilon^{t-1},H_t))\}
$$
is at most $\epsilon/K$. Consequently, the probability of the event $\cup_{i\leq K}{\cal E}_i$, which is exactly the event (\ref{eqevent}), is at most $\epsilon$, as claimed in (i).
\par
(ii): From the definition of $\Theta$ it follows that
$\|P_{t}E_t^Tg\|_1\leq \Theta \|g\|_1$
for every $g\in\bR^m$ and every $t$ and
$\upsilon^t\in \Upsilon^t$.
Now let $g\in \bR^m$ with $\|g\|_1=1$, and let $\gamma\geq0$.
We have
{\small$$
\begin{array}{l}
\bE_{t+1}\left\{\exp\{\gamma g^TS_{t+1}(\upsilon^{t+1})\}\right\}\\
= \bE_t^+\left\{\exp\{\gamma g^T S_t(\upsilon^t)\}\bE_{\zeta_{t+1}\sim P_{|\upsilon^{t},H_{t+1}}}\left\{\exp\{
[\gamma N^{-1}[E_{t+1}^Tg]^T[{\Phi}_{t+1}\left(H_{t+1}^T\beta\right)-\zeta_{t+1}]\}\right\}\right\}
\\
\leq \bE_t^+\left\{\exp\{\gamma g^T S_t(\upsilon^t)+\Psi(\|\gamma N^{-1}P_{t+1}E_{t+1}^Tg\|_1)\}\right\}
\hbox{\ [by (\ref{eq2}) and due to $\Psi_{t,\upsilon^{t-1},H_t}(\cdot)\leq\Psi(\cdot)$]}\\
\leq \bE_t^+\left\{\exp\{\gamma g^T S_t(\upsilon^t)\}\right\}\exp\{\Psi(\gamma \Theta /N)\},
\\
\multicolumn{1}{r}{\hbox{\ [since $\Psi$ is nondecreasing and $\|P_{t+1}E_{t+1}^Tg\|_1\leq\Theta\|g\|_1=\Theta$]}}
\\
= \bE_t\left\{\exp\{\gamma g^T S_t(\upsilon^t)\}\right\}\exp\{\Psi(\gamma \Theta /N)\},
\end{array}
$$}\noindent
whence
$$
\ln\left(\bE_{\upsilon^N}\left\{\gamma g^TS_N(\upsilon^N)\right\}\right)\leq N\Psi(\gamma \Theta /N).
$$
Therefore, for every $\delta>0$ one has, using Markov inequality
$$
\Prob\{g^TS_N(\upsilon^N)>\delta\}\leq \exp\{N\Psi(\gamma \Theta /N)-\gamma\delta\},
$$
whence finally
\begin{eqnarray}
\Prob\{g^TS_N(\upsilon^N)>\delta\} & \leq& \exp\{-N\Psi_*(\delta/\Theta)\},\\
\Psi_*(s) &=& \sup_{\alpha\geq0}\left[\alpha s-\Psi(\alpha)\right].
\end{eqnarray}
Taking into account that $G_{\upsilon^N}=S_N(\upsilon^N)$, selecting $g$ as $\pm$ basic orths in $\bR^m$ and using the union bound, we arrive at (\ref{thenprob}). (ii) is proved.
\qed

\subsection{Quality of parameter recovery}

Now we present a model parameter recovery performance bound,  as a consequence of the results in Section \ref{sec:concentration}.
Assume that we are under the premise of Proposition \ref{cor1}.ii and that the conditional by the past distributions of $\zeta_t$ are ``light tail'' ones, so that the worst case rate function $\Psi(\cdot)$ is such that $\Psi(s)\leq Cs$ for all $s\in[0,c]$, with properly selected $c>0$ and $C<\infty$. Specifying
\begin{equation}\label{specifying}
m=\kappa,\,\,E_{t}=\Lambda_t^{-1}(\upsilon^{t-1},H_t)H_{t},
\end{equation}
we get
$$
G_{\upsilon^N}=F_{\upsilon^N}(\beta).
$$
In this case, under the premise of Proposition \ref{cor1}.ii,  the typical $\|\cdot\|_\infty$ norm of $F_{\upsilon^N}(\beta)$ for large $N$ is of order of $\Theta\sqrt{\ln(\kappa)N}$. Thus, when there are reasons to
believe that $F_{\upsilon^N}$ is, typically,  strongly monotone on $\cB$ with parameter of strong monotonicity not deteriorating as $N$ grows,  we may hope that the solution to $\VI(F_{\upsilon^N},\cB)$ will be at the distance $O(N^{-1/2})$ from $\beta$.

To formulate the precise statement, let us define the $p$-modulus of strong monotonicity $\theta_p(\upsilon^N)$ of $F_{\upsilon^N}(\cdot)$ on $\cB$,  $p\in[1,\infty]$, as
$$
\theta_p(\upsilon^N)=\max\left\{\theta:\langle F_{\upsilon^N}(x)-F_{\upsilon^N}(y),x-y\rangle \geq \theta\|x-y\|_p^2,\,\,\forall x,y\in \cB\right\}.
$$
Note that the strict positivity of  $\theta_p(\upsilon^N)$ is independent of the value of $p$.
\begin{proposition}[Parameter recovery guarantee]\label{cor2}
Let $\upsilon^N$ be such that $\theta_p(\upsilon^N)>0$. Then the weak$\equiv$strong solution $\widehat{\beta}(\upsilon^N)$ to the variational inequality $\VI(F_{\upsilon^N},\cB)$ is unique, and
\begin{equation}\label{final}
\|\widehat{\beta}(\upsilon^N) - \beta \|_p\leq \|F_{\upsilon^N}(\beta)\|_\infty/\sqrt{\theta_p(\upsilon^N)\theta_1(\upsilon^N)}.
\end{equation}


\end{proposition}
\noindent{\bf Proof.}
We know that in the situation of Section \ref{GGLM_setup} weak solution $\widehat\beta=\widehat\beta(\upsilon^N)$ to $\VI(F_{\upsilon^N},\cB)$ exists and is a strong solution; besides this, it is well known that under the premise of Proposition, this solution is unique. Since $\widehat\beta$ is a strong solution, we have $\langle F_{\upsilon^N}(\widehat\beta), x - \widehat\beta \rangle \geq 0$ for all $x \in \cB$, and in particular $\langle F_{\upsilon^N}(\widehat\beta), \beta - \widehat\beta \rangle \geq 0$.
As a result,
$$
\|F_{\upsilon^N}(\beta)\|_\infty \|\beta-\widehat\beta \|_1\geq \langle F_{\upsilon^N}(\beta),\beta-\widehat\beta\rangle =
\underbrace{\langle F_{\upsilon^N}(\beta)-F_{\upsilon^N}(\widehat\beta),\beta-\widehat\beta\rangle}_{\geq\sqrt{\theta_p(\upsilon^N)\theta_1(\upsilon^N)}
\|\beta-\widehat\beta\|_p\|\beta-\widehat\beta\|_1}+\underbrace{\langle F_{\upsilon^N}(\widehat\beta),\beta-\widehat\beta\rangle}_{\geq0},
$$
and (\ref{final}) follows. \qed\par
To extract from Propositions \ref{cor1} information on performance guarantees of our estimate, we need to understand how to lower-bound the modulus of strong monotonicity. The related theoretical analysis would require heavy assumptions on inter-dependence of the subsequent ``regressors'' $H_{t}$, $t=1,2, \ldots$. This is an issue we do not want to touch here. Instead, let us focus on how to lower-bound  this modulus ``online.''  An immediate observation is that if $\Phi_t$ is strongly monotone, with positive moduli $\gamma_t$, on their domains:
$$
\langle \Phi_t(u)-\Phi_t(v),u-v\rangle \geq \gamma_t\langle u-v,u-v\rangle,\,\forall u,v\in\Dom\Phi_t,
$$
then
$$
\begin{array}{ll}
&\langle F_{\upsilon^N}(x)-F_{\upsilon^N}(y),x-y\rangle\\
 =&{1\over N}\sum_{t=1}^N\Lambda_t^{-1}(\upsilon^{t-1},H_t)\langle H_{t}\Phi_t(H_{t}^Tx)-
H_{t}\Phi_t(H_{t}^Ty),x-y\rangle\\
\geq& {1\over N}\sum_{t=1}^N\gamma_t\Lambda_t^{-1}(\upsilon^{t-1},H_t)\langle H_{t}^T (x-y),H_{t}^T(x-y)\rangle \\ =&(x-y)^T\underbrace{\left[{1\over N}{\sum}_{t=1}^N\gamma_t\Lambda_t^{-1}(\upsilon^{t-1},H_t)H_{t}H_{t}^T\right]}_{\Gamma(\upsilon^N)}(x-y),\\
\end{array}
$$
whence
\begin{equation}\label{eq122}
\theta_2(\upsilon^N) \geq \lambda_{\min}(\Gamma(\upsilon^N)),
\end{equation}
where $\lambda_{\min}(Q)$ is the minimal eigenvalue of a symmetric matrix $Q$. We see that $\theta_2(\upsilon^N)$ admits an ``online observable'' lower bound. We have also
$$
\theta_p(\upsilon^N)\geq \min_{h:\|h\|_p=1}h^T\Gamma(\upsilon^N)h=\left[\max_{g:\|g\|_{p_*}\leq1}g^T\Gamma^{-1}(\upsilon^N)g\right]^{-1}\eqno{[p_*={p\over p-1}]}.
$$
When $1\leq p\leq 2$, the right-hand side in this relation can be efficiently lower-bounded, the lower bound being tight within absolute constant \cite{NesterovQP} (equal to $\pi/2$ when $p=1$).
{\color{black}\subsubsection{The case of affine $\Phi_t$'s}\label{affinecase}
Assume that the links $\Phi_t(\cdot)$ are affine; in this situation, the vector field $F_{\upsilon^N}(\cdot)$ is affine as well. Assume also that the functions $\Psi_{t,\upsilon^{t-1},H_t}$ are observable. In this case observations $\upsilon^N$ imply $(1-\kappa\epsilon)$-reliable upper bound $\delta(\upsilon^N)$ on $\|F_{\upsilon^N}(\beta)\|_\infty$, specifically, the bound $\delta(\upsilon^N):=\max\limits_{1\leq j\leq\kappa}\delta_j(\upsilon^N)$ with $\delta_j(\upsilon^N)$ given by (\ref{eqevent}) as applied with
\[\myE_t=\Lambda^{-1}_t(\upsilon^{t-1},H_t)H_t.\] (Recall that this choice of $\myE_t$ results in $G_{\upsilon^N}=F_{\upsilon^N}(\beta)$). As a result, the observable under our assumptions convex set
\begin{equation}\label{myDelta}
\Delta(\upsilon^N)=\{x\in\cB:\|F_{\upsilon^N}(x)\|_\infty\leq\delta(\upsilon^N)\}
\end{equation}
is  ``$(1-\kappa\epsilon)$-confidence interval'' for $\beta$ -- this set contains $\beta$ with probability $\geq1-\kappa\epsilon$. In particular, given a norm $\|\cdot\|$ on $\bR^\kappa$, the quantity
$$
d(\upsilon^N)=\max\limits_{x\in \Delta(\upsilon^N)}\|x-\widehat{\beta}\|
$$
is an $(1-\kappa\epsilon)$-reliable upper bound on the $\|\cdot\|$-error of our estimate $\widehat{\beta}$. Whether this online error bound can or cannot be efficiently computed, depends on the norm $\|\cdot\|$; e.g., for the uniform norm $\|x\|_\infty=\max_i|x_i|$, 
%
computing the bound is easy - it reduces to $2\kappa$ maximizations of linear forms over $\Delta(\upsilon^N)$.\par
Note that under our present assumptions every policy $\cR$  for generating row vectors $\myE_t\in\bR^{1\times n_t}$ as deterministic functions of $\upsilon^{t-1},H_t$, $t=1,...,N$, induces an observable, given $\upsilon^N$, affine real-valued function
$$
G^{\cR}_{\upsilon^N}(x)={1\over N}\sum\limits_{t=1}^N \myE_t\left[\Phi_t(H_t^Tx)-\zeta_t\right]:\cB\to\bR
$$
along with  observable quantity $\delta^{\cR}_\epsilon(\upsilon^N)$, given by (\ref{eqevent}), such that
$$
\Prob\{|G^{\cR}_{\upsilon^N}(\beta)| \geq \delta^{\cR}(\upsilon^N)\}\leq\epsilon.
$$
Thus, the pair of observable linear inequalities $|G^{\cR}_{\upsilon^N}(\beta)|\leq\delta^{\cR}(\upsilon^N)$ on $\beta$  do hold true with probability $\geq 1-\epsilon$, and we can add these inequalities to the description of $\Delta(\upsilon^N))$ as given by (\ref{myDelta}), thus refining the confidence interval, and we can implement several refinements of this type stemming from several policies $\cR$. For an instructive application example, see Section \ref{instructive}.
}

\section{Illustrations: Spatio-Temporal Processes}\label{sec:app}
Assume that there are $L$ spatial locations; at discrete time $t$, the state of location $k\in\{1, \ldots, L\}$ is described by:
\begin{center}
realizations $\omega_{t,k}$ of random variable taking values in a set $Z\subset\bR^\mu$,
\end{center}
and $\mu$ is the number of possible states when the observation is non-zero.
Our observation at time $t$ is the block vector
$$
\zeta_t=[\omega_{t,1}; \ldots; \omega_{t,L}]\in\cZ= Z\times...\times Z\subset \mfZ=\bR^{\mu L}.
$$
We assume that the conditional (by what happened prior to time $t$) expectation of $\zeta_t$ depends solely on the collection
$$
(\zeta_{t-d},\zeta_{t-d+1}, \ldots, \zeta_{t-1})=\{\omega_{s,\ell}:t-d\leq s \leq t-1,1\leq \ell\leq L\},
$$
where $d\geq1$ is the ``memory depth'' of our process. Specifically, setting
$$
\zeta_\tau^t=(\zeta_\tau,\zeta_{\tau+1}, \ldots, \zeta_t)=\{\omega_{s,\ell}\in\bR^\mu: \tau\leq s\leq t,1\leq\ell\leq L\}, \zeta^t=\zeta_{-d+1}^t.
$$
Assume that our observations start at time $-d+1$ and that for  every $t\geq 1$ the conditional, given $\zeta^{t-1}$, expectation $\bE_{|\zeta^{t-1}}$ of $\zeta_t$ is
\begin{equation}\label{tomeet}
\bE_{|\zeta^{t-1}}\{\zeta_t\}=\Phi(\eta^T(\zeta_{t-d}^{t-1})\beta),
\end{equation}
where
\begin{itemize}
\item
$\beta$ is the vector of parameters of the process. This vector is obtained by  writing down in once for every prescribed order the entries of the elements of the collection
$$
\{\beta^0\in\bR^{\mu L},\beta^s\in\bR^{\mu L\times\mu L},1\leq s\leq d\}
$$
of ``actual parameters,'' and
$$
\eta^T(\zeta_{t-d}^{t-1})\beta=\beta^0+\sum_{s=1}^b\beta^s\zeta_{t-1}\in\bR^{\mu L}.
$$
In other words, denoting by
$$
\kappa=\mu L+d\mu^2L^2
$$
the dimension of $\beta$, $\eta(\gamma)$ is matrix-valued function, taking values in $\bR^{\kappa\times \mu L}$, of $L\times d$ array $\xi=\{\xi_{s\ell},1\leq s\leq d,1\leq\ell\leq L\}$ of vectors $\xi_{s\ell}\in\bR^\mu$, and this function is uniquely defined by the requirement that
$$
\forall x\in\mfB: \eta^T(\xi)x=x^0+\sum_{s=1}^dx^s[\xi_{t-s,1}; \ldots ;\xi_{t-s,L}],
$$
where $\mfB=\bR^\kappa$ is the space of column vectors $x$ obtained by arranging in a prescribed order the entries in a collection
$\{x^0\in\bR^{\mu L},x^s\in\bR^{\mu L\times\mu L},1\leq s\leq d\}$.
\par
We always assume that $\beta$ resides in a known in advance convex compact subset $\cB$ of $\mfB$;
\item $\Phi(\cdot):\Dom\Phi\to\mfZ$ is the link function -- a continuous and monotone vector field on a closed convex domain   $\Dom\Phi\subset\mfZ$. We always assume that whenever $\xi\in\cZ\times\ldots\times\cZ$
and $x\in\cB$, we have $\eta^T(\xi)x\in\Dom\Phi.$
\end{itemize}
A spatio-temporal process we just have described can be modeled by stationary GGLM, where matrix component $H_t$ of observations at time $t$ is known deterministic function of $\zeta_{t-d},\zeta_{t-d+1}, \ldots, \zeta_{t-1}$:
$$
H_t=\eta(\zeta_{t-d}^{t-1}) \in \bR^{\kappa \times \mu L},
$$
and all other components of the model are readily given by the above description. Consequently, we can apply the machinery from Section \ref{sec:paramrec} to recover the parameters $\beta$ from observations $\zeta_t$ on time horizon $-d+1\leq t\leq N$.
\par
We are especially interested in the case of {\sl categorial} and {\sl Poisson} spatio-temporal processes.

\subsection{Categorial spatio-temporal process}\label{sec:ber}

The spatio-temporal categorial process with $\mu+1$ states models the situation when at time $t$ the $k$-th location can be in one of $\mu+1$ states --- {\sl ground state} 0 and {\sl nontrivial states} $1, \ldots, \mu$. We specify the model as follows:
\begin{itemize}
\item encode these states by vectors from $\bR^\mu$, with the ground state encoded by the origin and nontrivial state $i$ encoded by $i$-th basic orth in $\bR^\mu$. As a result, $Z$ becomes the finite set -- the vertices of the simplex $\{z\in\bR^\mu_+:\sum_iz_i\leq 1\}$, $\cZ$ becomes the set of Boolean block-vectors with $L$ blocks of dimension $\mu$ each, and at most one nonzero entry in every block. A special case is when $\mu = 1$, then the model becomes the Bernoulli process where the observations are binary vectors.
\item assume that the states $\omega_{tk}$ are conditionally on $\zeta^{t-1}$ independent across $k$.
\item the values of $\Phi$  are block vectors with $L$ blocks of dimension $\mu$ each; denoting by $\Phi_{ik}(z)$ the $i$-th entry of $k$-th block in $\Phi(z)$.
\end{itemize}
In order to meet (\ref{tomeet}), the quantity $\Phi_{ik}(\eta^T(\zeta_{t-d}^{t-1})\beta)$ should be the conditional on $\zeta^{t-1}$ probability for $k$-th location at time $t$ to be at active state $i$. Thus, it should be nonnegative, and the sum of these quantities should be $\leq 1$. Consequently, when speaking about categorial spatio-temporal processes, we always assume that we are given a closed convex set $\overline{Z}\subset \Dom\Phi$ such that
$$
\forall z\in\overline{Z}: \Phi(z)\geq0\ \& \sum_{i=1}^\mu\Phi_{ik}(z)\leq 1,\, 1\leq k\leq L,
$$
and that
$$
\eta^T(\zeta_1^d)x\in\overline{Z}\,\,\forall x\in\cB, \zeta_1^d \in\cZ\times\cdots\times \cZ.
$$
Note that the conditional on $\zeta^{t-1}$ probability for $k$-th location at time $t$ to be in ground state is
$$
1-\sum_{i=1}^{\mu} \Phi_{ik}(\eta^T(\zeta_{t-d}^{t-1})\beta).
$$
Categorial spatio-temporal process of the outlined type and VI-based techniques for parameter recovery in these processes were considered in \cite{juditsky2020convex}.
\par
Observe that in the resulting stationary GGLM for every $t\geq1$ and $h\in\mfZ$ setting
\begin{center}
$\nu=\nu_t(\upsilon^{t-1},H_t)=\Phi(H_{t}^T\beta)$, we get
$
\nu\geq0,\|\nu\|_\infty\leq1,
$
\end{center}
and
\begin{equation}\label{april24}
\begin{array}{rcl}
\bE_{|\upsilon^{t-1},H_t}\left\{\exp\{h^T\zeta_t\}\right\}&=&
\prod_{k=1}^L \left( 1+\sum_{i=1}^\mu \nu_{ik}[{\rm e}^{h_{ik}}-1]  \right),
\\
\multicolumn{1}{l}{\Rightarrow}&&\\
\ln\left(\bE_{|\upsilon^{t-1},H_t}\left\{\exp\{h^T\zeta_t\}\right\}\right)&=&h^T\nu+
\left[\sum_{k=1}^L\ln\left(1+\sum_{i=1}^\mu\nu_{ik}[{\rm e}^{h_{ik}}-1]\right)-h^T\nu\right].
\end{array}
\end{equation}
The function
$$
f_\nu(g)=\sum_{k=1}^L\ln\left(1+\sum_{i=1}^\mu\nu_{ik}[{\rm e}^{g_{ik}}-1]\right)-g^T\nu
$$
of $g\in\mfZ$ is convex, and therefore its maximum over the set $\{g:\|g\|_1\leq r:=\|h\|_1\}$ is achieved at a vertex,  where all but one entry of $g$ are zero, and the remaining entry is $\pm r$.\footnote{The unit $\ell_1$ ball is also known as the cross-polytope; the vertices of a cross-polytope can be the canonical basis, i.e., all the vectors as permutations of $(\pm 1, 0, 0, \ldots, 0)$.} As a result, setting
\def\cat{{\mathrm{\tiny cat}}}
{\footnotesize$$
\begin{array}{l}
\alpha_t(\upsilon^{t-1},H_t)=\max\limits_{x\in\cB} \|\Phi(H_{t}^Tx)\|_\infty\in[\|\nu_t\|_\infty,1],\\
\Psi^\cat_{t,\upsilon^{t-1},H_t}(s):=\max\limits_{\alpha}\left\{\max\left[\ln\left(1+\alpha[{\rm e}^s-1]\right)-\alpha s,\ln\left(1+\alpha[{\rm e}^{-s}-1]\right)+\alpha s\right]:0\leq\alpha\leq\alpha_t(\upsilon^{t-1},H_t)\right\}
\\
\leq \Psi^\cat(s):=\max\limits_{\alpha}\left\{\max\left[\ln\left(1+\alpha[{\rm e}^s-1]\right)-\alpha s,\ln\left(1+\alpha[{\rm e}^{-s}-1]\right)+\alpha s\right]:0\leq\alpha\leq1\right\}\\
\end{array}
$$}\noindent
we get continuous convex and even function of $s\in\bR$, with $\Psi_{t,\upsilon^{t-1},H_t}(\cdot)$ observable at time $t$,  such that
$$
\forall (h\in\mfZ,t\geq1,\upsilon_s\in\Upsilon_s,s\leq t): \bE_{|\upsilon^{t-1},H_t}\left\{\exp\{h^T\zeta_t\}\right\}\leq h^T\Phi(H_{t}^T\beta)+\Psi\yx{_{t,\upsilon^{t-1},H_t}}{}^\cat(\|h\|_1),
$$
that is, relation (\ref{eq2}) holds true with  $\Psi_{t,\upsilon^{t-1},H_t}=\Psi^\cat_{t,\upsilon^{t-1},H_t}$ and the identity matrix in the role of $P_t$.
\par
Observe also that,  as it is immediately seen, whenever $\zeta_s$, $1\leq s\leq d$, are Boolean vectors (which definitely is the case when $\zeta_s\in\cZ$), $s\leq d$, the matrix $\eta^T(\yx{\zeta_1,...,\zeta_d}{\zeta_1^d})$ is Boolean, and every column of it has at most one nonzero entry, that is, the columns in all realizations of $H_{t}$ are of $\ell_1$-norm not exceeding 1. As a result, when applying Proposition \ref{cor1} to a spatio-temporal categorial process, we can use  the $\mu L\times \mu L$ identity matrix in the role of $P_t$'s, and set $\Psi_{t,\upsilon^{t-1},H_t}\equiv \Psi^\cat_{t,\upsilon^{t-1},H_t}$. Further, when specifying $E_{t}$ according to (\ref{specifying}), resulting in $G_{\upsilon^N}=F_{\upsilon^N}$, we can set $\Theta_{tj}(\upsilon^{t-1},H_t)=\Lambda_t^{-1}(\upsilon^{t-1},H_t)$ in (i); this choice works also whenever the $\|\cdot\|_1$-norms of all columns in $E_{t}$ are bounded by 1. When, in addition, $\Lambda_t(\cdot)\equiv 1$, the premise of item (ii) is satisfied with $\Theta=1$ and $\Psi=\Psi^\cat$.

\subsection{Poisson spatio-temporal process}\label{sec:Poisson}

Spatio-temporal Poisson process models the situation where the state $\omega_{tk}$ of the $k$-th location is a non-negative integer (that is, $\mu=1$ and $Z=\{0,1,2, \ldots \}$), and the conditional (by the past) distribution of $\zeta_t=[\omega_{t1};\omega_{t2}; \ldots ;\omega_{tL}]$ is $k$ independent Poisson components $\omega_{tk}$ with parameters specified by $\Phi_k(\eta^T(\zeta_{t-d}^{t-1})\beta)$. In order to meet (\ref{tomeet}) such that the expectation of the Poisson random variable is nonnegative, we should assume now that we are given a closed  convex set $\overline{Z}\subset \Dom\Phi$ such that
$$
\Phi(z)\geq0\,\forall z\in\overline{Z}
$$
and that for all nonnegative integral vectors $z_s\in\bR^L$, $1\leq s\leq d$, it holds
$$
\eta^T(z_1, \ldots ,z_d)x\in\overline{Z}\,\,\forall x\in\cB.
$$

Observe that in the resulting stationary GGLM  for all $t\geq1$ and all $h=[h_1; \ldots ; h_L]\in\mfZ$, setting $\nu_t=\Phi(H_{t}^T\beta)$, based on the conditional independence across locations, and applying the Poisson moment generating function  for each location $$\bE\{\exp\{t\xi\} \}= \exp\{\lambda e^t -1\}, \quad \xi \sim \mbox{Poisson}(\lambda),$$ we have
\begin{equation}\label{april24a}
\begin{array}{rcl}
\bE_{|\upsilon^{t-1},H_t}\left\{\exp\{h^T\zeta_t\}\right\}&=&
\exp\left\{\sum_{k=1}^L[\nu_t]_k(\exp\{h_k\}-1)\right\},\\
\multicolumn{1}{l}{\Rightarrow}&&\\
\ln\left(\bE_{|\upsilon^{t-1},H_t}\left\{\exp\{h^T\zeta_t\}\right\}\right)&=&h^T\nu_t+
\sum_{k=1}^L[\nu_t]_k\left[\exp\{h_k\}-h_k-1\right]\\
&=&h^T\Phi(H_{t}^T\beta) +\sum_{k=1}^L[\nu_t]_{k}\left[\exp\{h_{k}\}-h_{k}-1\right].\\
\end{array}
\end{equation}
Now let us set
$$
\chi_t(\upsilon^{t-1},H_t)=\max_{u\in\cB}\|\Phi(H^T_{t}u)\|_\infty,\,\,f(s)=\exp\{s\}-s-1.
$$
Then $\|\nu_t\|_\infty\leq\chi_t(\upsilon^{t-1},H_t)$ and
$$
\begin{array}{l}
\sum_{k=1}^L[\nu_t]_{k}\left[\exp\{h_{k}\}-h_{k}-1\right]\\
\leq \|\nu_t\|_\infty\|[f(h_1); f(h_2); \ldots; f(h_L)]\|_1\leq
\chi_t(\upsilon^{t-1}, H_t)\max_g\left\{E(g):\|g\|_1\leq\|h\|_1\right\},\\
\end{array}
$$
where $E(g)=\|[f(g_1);...;f(g_L)]\|_1$.
Since $f$ is nonnegative and convex on $\bR$, the function $E(g)$ is convex, and therefore its maximum over the $\ell_1$-ball $\{g:\|g\|_1\leq\|h\|_1\}$ is achieved at a vertex, implying that
$$
\max_g\{E(g):\|g\|_1\leq\|h\|_1\}=f(\|h\|_1).
$$
Thus, setting
$$
\Psi_{t,\upsilon^{t-1},H_t}(r)=\chi_t(\upsilon^{t-1},H_t)\Psi(r),\,\Psi(r):=\exp\{r\}-r-1, \,\,P_t=I_L,
$$
and invoking (\ref{april24a}), we conclude that
$$
\bE_{|\upsilon^{t-1},H_t}\left\{\exp\{h^T\zeta_t\}\right\}\leq  h^T\Phi_t(H_{t}^T\beta)+
\Psi_{t,\upsilon^{t-1}, H_t}(\|P_th\|_1),
$$
as required by (\ref{eq2}). Note that $\Psi_{t,\upsilon^{t-1},H_t}(\cdot)$, is, modulo computational aspects, observable at time $t$.

\subsection{Numerical illustration for Poisson case}\label{instructive}

Now we present some numerical examples of the Poisson spatio-temporal process case to demonstrate the computational procedure for estimation and online bounds for the recovery error. In the experiments to be reported, we dealt with the identity link function (thus, the non-negativity of the Poisson intensity is ensured by constraints on the feasible parameters.) The process was
$$
[\zeta_t]_k\sim\mathrm{Poisson}\left(\beta^0_k+\sum_{s=1}^d\sum_{\ell=1}^L\beta^s_{k\ell}[\zeta_{t-s}]_\ell\right),\,t\geq1, 1\leq k\leq L.
$$
Recall that $d$ is the memory depth, $L$ is the number of locations, and $[\zeta_t]_k$ is the $k$-th coordinate location of response at time $t$.
\paragraph{Generating parameter $\beta$.} Since $\Phi$ is the identity,  selection of $\beta$ should ensure nonnegativity of the quantities
$$
\beta^0_k+\sum_{s=1}^d\sum_{\ell=1}^L\beta^s_{k\ell}[\zeta_{t-s}]_\ell,\,k\leq L,
$$
whatever be nonnegative integral $\zeta_s\in\bR^L$. 
To this end, we restrict $\beta$ to be nonnegative. Besides this, with positive $\beta^0_i$, in order for the process not to explode, we should have
$$
\sum_{s=1}^d\sum_{\ell=1}^L\beta^s_{k\ell}<1,\,k\leq L.
$$
Our generation was governed by two parameters $a>0$ and $b\leq 1$, specifically, we ensured that
\begin{equation}\label{generation}
\beta\geq0\ \&\ \beta^0_k=a\ \&\ \sum_{s=1}^d\sum_{\ell=1}^L\beta^s_{k\ell}=b,\,\,k\leq L.
\end{equation}
Generation was organized as follows: given $d$, $L$, we selected 
$\beta$ satisfying (\ref{generation}) at random
\par The set $\cB$ -- a priori localized for $\beta$ -- also was given by parameters $a,b$ according to
$$
\cB=\{x=\{x^0_k,x^s_{k\ell},1\leq s\leq d,1\leq k,\ell\leq L\}: x \geq 0,x^0_k\leq 1.1a,\sum_{s=1}^d \sum_{\ell=1}^Lx^s_{k\ell}\leq 1.1 b\,\forall k.\}
$$
\paragraph{Building estimate $\widehat{\beta}$.} Our estimate was exactly as explained in the special case of Section \ref{sec:paramrec} with $\Lambda_t\equiv 1$, that is, $\widehat{\beta}(\upsilon^N)$ was an optimal solution to the Least Squares problem
$$
\min\limits_{x\in\cB}\left[{1\over 2N}\sum_{t=1}^N\|H_t^Tx-\zeta_t\|_2^2:={1\over 2N}\sum_{t=1}^N\sum_{k=1}^L\left([\zeta_t]_k-x^0_k-\sum_{s=1}^d\sum_{\ell=1}^Lx^s_{k\ell}[\zeta_{t-s}]_\ell\right)^2\right].
$$
\paragraph{On-line bounds.} The major goal of our experimentation was to get online upper bounds on the recovery errors.  As it turned out, bounds stemming from Proposition \ref{cor2} under the circumstances were quite poor -- the values of the resulting error bounds typically were larger than the corresponding sizes of our a priori parameter's localizer $\cB$. We used the approach described in Section \ref{affinecase} to get meaningful error bounds. The choice of $M$ and $\alpha_i$'s will be explained later.
Specifically, we
\begin{enumerate}
\item selected $M$ of policies  $\cR^\iota$, $1\leq\iota\leq M$, of generating $L$-dimensional row vectors $\myE_t$ as deterministic functions of $\upsilon^{t-1}$ and $H_t$, resulting in $M$ affine functions
$$
G^\iota_{\upsilon^N}(x)={1\over N}\sum_{t=1}^N\myE_t^\iota[H_tx- \zeta_t];
$$
($\myE_t^\iota$ are yielded by $\cR^\iota$.)
\item used Proposition \ref{cor1}.i and the preceding results of this section to build online $(1-\epsilon/M)$-reliable upper bounds on the quantities $|G^\iota_{\upsilon^N}(\beta)|$, specifically, the bounds
\begin{equation}\label{bounds}
\begin{array}{c}
\delta^\iota(\upsilon^N)=\min_i\left[\alpha_i\ln(2KM/\epsilon)+ \alpha_i \sum_{t=1}^N\Psi_t(\alpha_i^{-1}N^{-1}\|\myE^\iota_t\|_1)\right]\\
\Psi_t(r)=\chi_t[\exp\{r\}-r-1],\\
\chi_t=\max\limits_{x\in\cB}\|H_t^Tx\|_\infty,
\end{array}
\end{equation}
on $|G^\iota_{\upsilon^N}(\beta)|$.
\item specified $(1-\epsilon)$-confidence interval
$$
\Delta(\upsilon^N)=\{x\in\cB: |G^\iota_{\upsilon^N}(x)|\leq\delta^\iota(\upsilon^N), 1\leq \iota\leq M\}
$$
-- under the circumstances, this is a convex compact set that contains $\beta$ with probability at least $1-\epsilon$.
\item built induced confidence intervals $\Delta^0_k$, $\Delta^s_{k\ell}$ on the entries in $\beta$, the endpoints of these intervals being the minima and the maxima of the corresponding entries in $x$ taken over $x\in\Delta(\upsilon^N)$. \\
By construction, the probability for all entries in $\beta$ to be covered by the corresponding intervals is at least $1-\epsilon$, so that the maximal, over points from the intervals, deviations of the points from the corresponding entries in $\widehat{\beta}$ form a vector which is an $(1-\epsilon)$-reliable upper bound on the vector of magnitudes of the recovery errors.
\end{enumerate}
The essence of the matter is how to specify the policies $\cR^\iota$. The first $\kappa$ of our policies resulting in {\sl basic} online bounds were to take, as  $\myE^\iota_t$, the $\iota$'th row in $H_t$, resulting in
$$[G^1_{\upsilon^N}(x); \ldots ;G^\kappa_{\upsilon^N}(x)]=F_{\upsilon^N}(x).$$
The remaining policies were more sophisticated and inspired by the following consideration.
\begin{quote}
Given index $k\leq\kappa$ of an entry in $\beta$ and a ``scale parameter'' $\theta>0$, let us impose on $\myE_t^\iota$'s the restriction $\|\myE_t^\iota\|_1\leq\theta$, thus imposing upper bounds $\Psi_t(\alpha_i^{-1}N^{-1}\theta)$ on the terms $\Psi_t(\alpha_i^{-1}N^{-1}\|\myE^\iota_t\|_1)$ in (\ref{bounds}), and under this assumption let us try to select $\myE_t^\iota$'s in order to make the slope of the associated affine function $G_{\upsilon^N}^\iota(\cdot)$ as close as possible to $k$-th basic orth in $\bR^\kappa$. Assume for a moment that we fully succeeded and the slope is exactly this basic orth. Then we get at our disposal inequality an inequality for $s$:
$$ |s-c|\leq\delta=\delta^\iota(\upsilon^N),\,c:={1\over N}\sum_{t=1}^N\myE_t^\iota H_t\zeta_t,$$ in variable $s$ with hopefully small $\delta$ with is with overwhelming probability satisfied by $k$-th entry of $\beta$.
\par
The outlined methodology reduces to finding  $L$-dimensional row vectors $\myE_t$ of $\|\cdot\|_1$-norm not exdceeding  a given $\theta$ such that the vector
$$
{1\over N}\sum_{t=1}^TH_t\myE_t^T
$$
is as close as possible to $k$-th basic orth $e_k$ in $\bR^\kappa$. The difficulty here is that $\myE_t$ should be ``non-anticipating'' -- they should be specified in terms of $\nu^{t-1}$ and $H_t$. The simplest policy of this type is the advanced policy
\def\Argmin{\mathrm{Argmin}}
$$
\begin{array}{rcl}
f_1&=&e_k,\\
\myE_t&\in&\Argmin_g\left\{\|f_t-N^{-1}H_tg\|:\|g\|_1\leq\theta\right\},  \\
f_{t+1}&=&f_t-N^{-1}\myE_t.\\
\end{array}
$$
In our experiments, the only one of the standard norms which worked in this advanced policy was $\|\cdot\|_2$, and these were the policies we used. To implement such a policy, we need to specify $\theta$. This parameter is responsible for the tradeoff between the closeness of $e_k(\upsilon^N):={1\over N}\sum_{t=1}^TH_t\myE_t^T$ to our target $e_k$ and online upper bound
$$
\delta=\min_i\left[\alpha_i\ln(2KM/\epsilon)+\sum_{t=1}^N\alpha_i\Psi_t(\alpha_i^{-1}N^{-1}\theta)\right]
$$
on $|e_k^T(\upsilon^n)\beta-{1\over N}\sum_{t=1}^N\myE_tH_t\zeta_t|$ we end up with. Resolving this tradeoff analytically seems to be an impossible task, but we can use the same computation-oriented approach as with optimizing in the scale parameter $\alpha$, specifically, select in advance a finite grid $\Theta=\{\theta_k:k\leq K\}$ of values of $\theta$  and run $K$ advanced policies parameterized by $\theta\in\Theta$.
\end{quote}
\par
\paragraph{``Advanced bounds.''} In our implementation, we selected a $K$-element set $\Theta$ on the positive ray and, on top of already explained $\kappa$ policies $\cR^\iota$, $\iota\leq \kappa$, given by $\myE_t=\mathrm{Row}_\iota(H_t)$ used $\kappa K$ policies more, associating with every $k\leq\kappa$, the $K$ aforementioned advanced policies with common target $e_k$ and different values of $\theta$  running through $\Theta$.
\par
\begin{figure}[htbp]
    \centering
    \begin{minipage}{0.49\textwidth}
    \begin{center}
        \includegraphics[height=155pt,width=200pt]{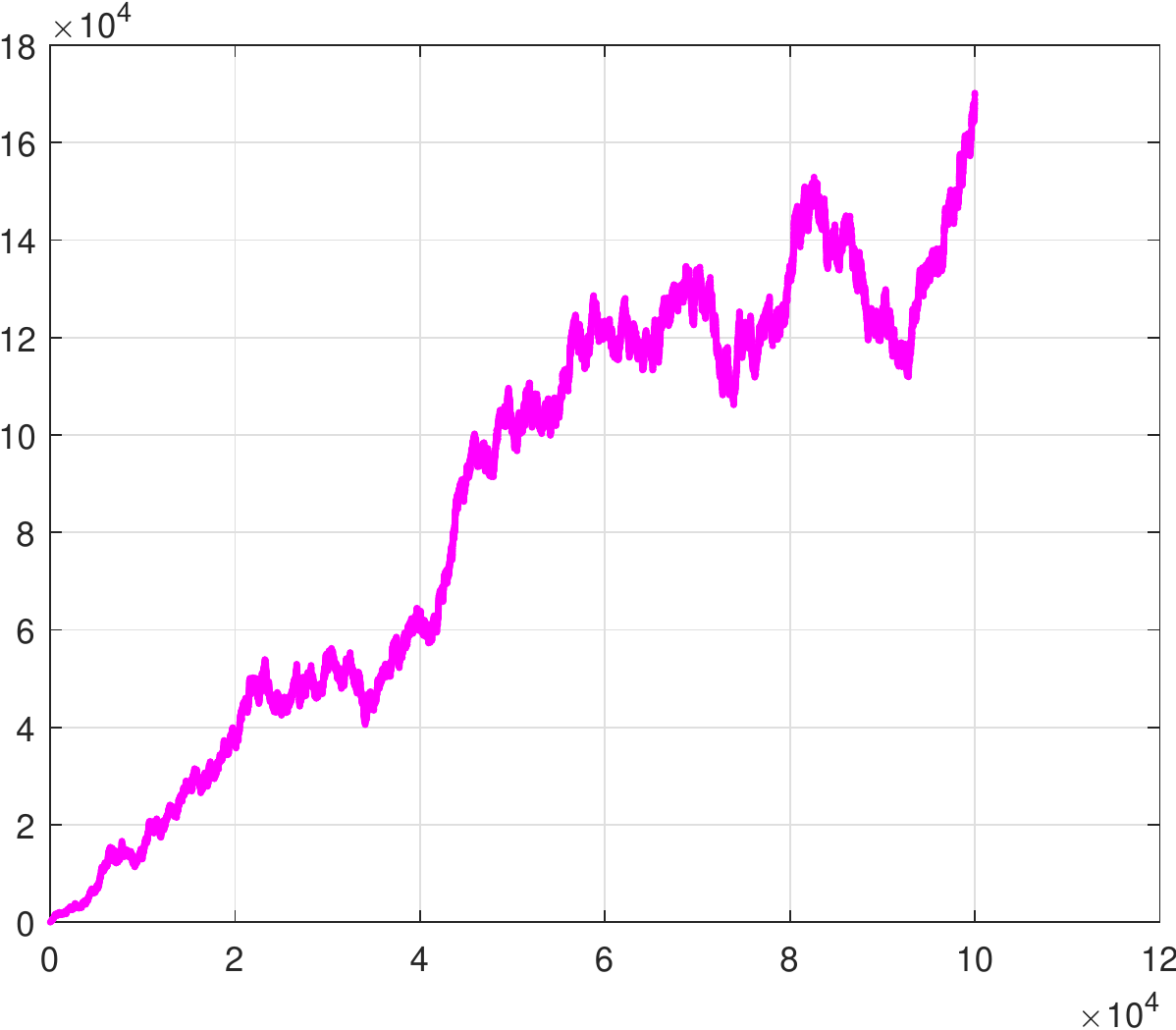}
        \subcaption{Experiment A, ${1\over LN}\sum_{t}\omega^{\hbox{\tiny total}}_t\approx$1.7e4}
        \end{center}
    \end{minipage}
    \begin{minipage}{0.49\textwidth}
     \begin{center}
        \includegraphics[height=150pt,width=200pt]{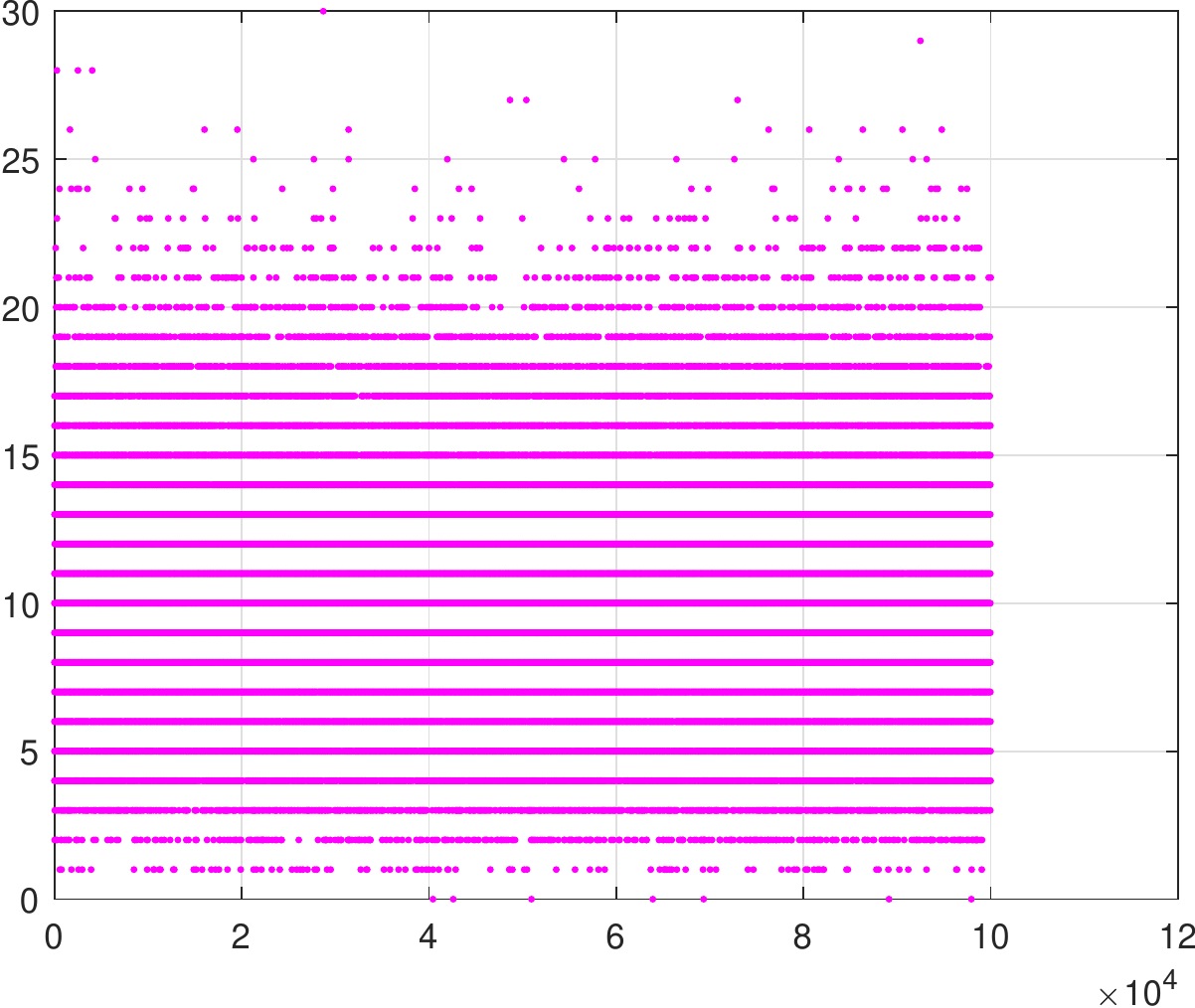}
        \subcaption{Experiment B, ${1\over LN}\sum_t\omega^{\hbox{\tiny total}}_t=2.00$}
        \end{center}
    \end{minipage}
    \caption{Sample trajectories of $\omega^{\hbox{\tiny total}}_t$ versus $t$.}
    \label{Fig:Trajectory}
\end{figure}

\paragraph{Implementation and numerical results.} In the experiments to be reported, we used $d=L=5$, resulting in $\kappa=130$, $N=100,000$, $\epsilon=0.01$, and $M=780=5\times 130+130$ - according to the above explanation, $M$ is $\kappa$ plus $\kappa$ times the cardinality of $\Theta$.
$$
\Gamma=\{\alpha_i:=10^{0.25 i-4},0\leq i\leq 36\}, \,\,\Theta=\{0.5,0.75,1.00,1.25,2\}.
$$
\textbf{Numerical results.} We present numerical results for two (in fact, quite representative) experiments. In the first (Experiment A) we used $a=1$, $b=1$, resulting in self-exciting process with growing with time $t$ total, over locations, number $\omega^{\hbox{\tiny total}}_t:=\sum_{k=1}^L\omega_{tk}$ of events
at time $t$. In the second (Experiment B), we used $a=1$, $b=0.5$, resulting in a stable process, as shown in Fig. \ref{Fig:Trajectory}.

\par
\paragraph{Recovery errors.} Now, we report the graphical and numerical data on actual recovery errors and online bounds on these errors. On the plots and in the tables, ``basic'' online bounds stem from the $\kappa$ initial policies $\cR^\iota$, $\iota\leq \kappa$, while ``advanced'' bounds stem from all $\kappa(K+1)$ policies described above.

\begin{figure}[!t]
 \begin{center}
    \begin{minipage}{0.49\textwidth}
    \begin{center}
        \includegraphics[height=150pt,width=200pt]{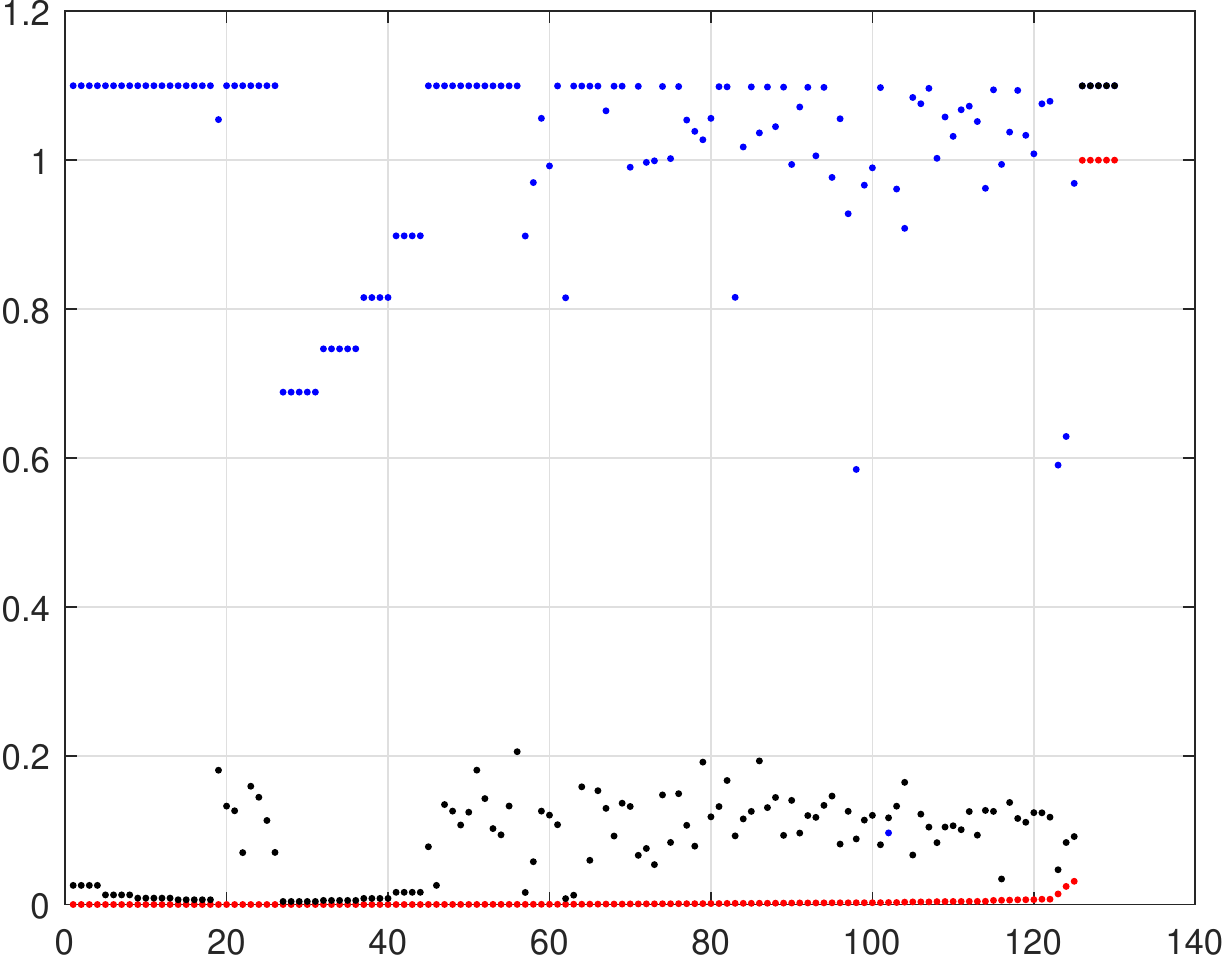}
        \subcaption{Experiment A}
        \end{center}
    \end{minipage}
    \begin{minipage}{0.49\textwidth}
    \begin{center}
        \includegraphics[height=150pt,width=200pt]{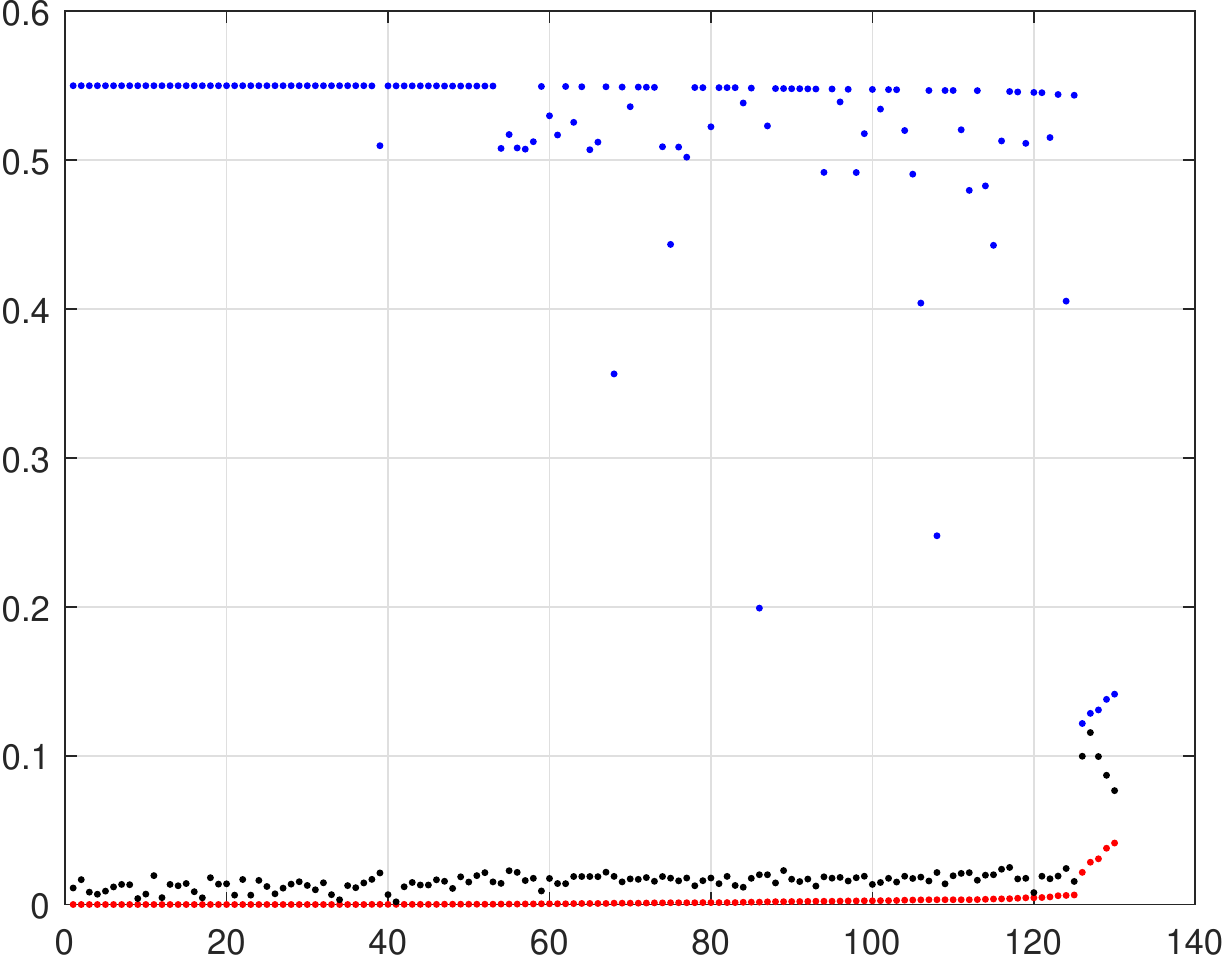}
        \subcaption{Experiment B}
        \end{center}
    \end{minipage}
    \end{center}
    \caption{Coordinate-wise recovery errors and online upper bounds on these errors.
 Red dots: actual coordinate-wise recovery errors; blue dots: basic online bounds; black dots: advanced online bounds.
Ordering of coordinates makes the errors non-decreasing.}
\label{Fig:Coordinate-error}
\end{figure}


\begin{table}[!b]
\centering
\caption{Recovery error and bounds for simulated examples.}
\begin{subtable}{\linewidth}\centering
\resizebox{0.75\linewidth}{!}{
\begin{tabular}{|c|c|c|c|c|}
     \cline{2-5}
    \multicolumn{1}{c|}{}&$\min$&\hbox{median}&\hbox{mean}&$\max$\\
    \hline
    \hbox{actual errors}&0.0000/0.0000&0.0005/0.0004&0.0402/0.0018&1.0000/0.0312\\
    \hline
    \hbox{bounds, basic}&0.0962/0.0962&1.0775/1.0725&1.0045/1.0007&1.100/1.100\\
    \hline
    \hbox{bounds, advanced}&0.0041/0.0041&0.0983/0.0932&0.1215/0.0824&1.100/0.2053\\
\hline
\end{tabular}
}
\caption{Experiment A, coordinate-wise errors and their online bounds}
\end{subtable}

\begin{subtable}{\linewidth}\centering
\resizebox{0.75\linewidth}{!}{
\begin{tabular}{|c|c|c|c|c|}
     \cline{2-5}
\multicolumn{1}{c|}{}&$\min$&\hbox{median}&\hbox{mean}&$\max$\\
\hline
\hbox{actual errors}&0.0000/0.0000&0.0007/0.0006&0.0024/0.0013&0.0413/0.0064\\
\hline
\hbox{bounds, basic}&0.1216/0.1991&0.5485/0.5487&0.5140/0.5293&0.5500/0.5500\\
\hline
\hbox{bounds, advanced}&0.0018/0.0018&0.0161/0.0158&0.0183/0.0152&0.1155/0.0249\\
\hline
\end{tabular}
}
\caption{Experiment B, coordinate-wise errors and their online bounds}
\end{subtable}
\label{Table1}
\end{table}


\noindent

In Table \ref{Table1}, the first number in a cell represents the ``column quantity,'' say, the median, taken over all $\kappa=130$ coordinate-wise errors and their upper bounds, while the second number corresponds to recovery errors/bounds on the errors for 125
``location to location influence'' parameters $\beta^s_{k\ell}$. This distinction is because in Experiment A, the ``birth rates'' $\beta^0_k$ do not admit nontrivial estimates. Indeed, with our setup, these birth rates are equal to 1, while the typical observation in the unstable case is of the order of 10$^4$; of course, there is absolutely no way to recover birthrates of the order of 1 on the ``background'' of magnitude 10$^4$. In Experiment A, the estimates of birth rates were nearly zero; this quite natural fact is responsible for the ``large'' -- equal to 1 -- uniform recovery error in observed in this experiment. As the corresponding table shows, the uniform norm of recovering all but 5 ``birth rate'' coefficients in $\beta$ is 0.03 - 30 times smaller than the uniform norm of recovering the entire $\beta$. A similar albeit less profound phenomenon takes place in Experiment B.
\par
From the above data, it is clear that utilizing advanced 
policies improved significantly the quality of online error bounds. While even after this improvement, the advanced bounds are much larger than the actual errors, these bounds seem to yield some information.
\paragraph{Upper-bounding $\|F_{\upsilon^N}(\beta)\|_\infty$.} Our experiments show that the online upper bound on $\|F_{\upsilon^N}(\beta)\|_\infty$ is within factor 5-10 of the actual value of the quantity.
\paragraph{Predictive power.} One way to utilize an estimate $\widehat{\beta}$ of $\beta$ is to use the observations obtained so far to predict the states in the locations in the future. Specifically, given observations $\omega^t:=\{\omega_{\tau k}:\tau\leq t,k\leq L\}$, and prediction step  $p\geq1$, we can write a simple recurrence specifying the conditional, $\omega^t$ given, expectations $\bar{\omega}_{t+p,k}$ of the states $\omega_{t+p,k}$ and use these expectations to build confidence intervals for $\omega_{t+p,k}$ {\sl as if} the conditional, $\omega^t$ given, distributions of these random variables were $\mathrm{Poisson}(\bar{\omega}_{t+p,k})$ (``as if'' reflects the fact that the actual conditional distributions in question probably are $\mathrm{Poisson}(\bar{\omega}_{t+p,k})$ only when $p=1$). To write down the aforementioned recurrence, we need to know the process's true parameters $\beta$; by replacing these parameters with their estimates $\widehat{\beta}$, we get ``empirical'' confidence intervals for future states. Table \ref{Table:experimentB} shows how the prediction worked in the testing sample.

\begin{table}[!t]
\centering
\caption{Average, over 100,000-element testing sample, frequency for $p$-step-ahead
empirical 0.95-confidence intervals
\underline{not} to cover the actual states of locations $k$,   $1\leq k\leq 5$.
 Last row: trivial prediction with $\bar{\omega}_{t+p,k}$ set to $\omega_{tk}$ (``tomorrow will be same as today'').
 }
\begin{subtable}{0.49\linewidth}\centering
\resizebox{0.8\linewidth}{!}{
\begin{tabular}{|c|c|c|c|c|c|}
    \hline
$p$&$k=1$&$k=2$&$k=3$&$k=4$&$k=5$\\
\hline\hline
    1&  0.049&0.050&0.050&0.048&0.050\\ \hline
       2&  0.049&0.051&0.050&0.050&0.050\\ \hline
       3&  0.088&0.054&0.054&0.054&0.068\\ \hline
       4&  0.088&0.056&0.057&0.058&0.113\\ \hline
       5&  0.135&0.063&0.063&0.061&0.116\\ \hline
       6&  0.137&0.069&0.068&0.080&0.142\\ \hline
       7&  0.166&0.075&0.070&0.084&0.162\\ \hline
       8&  0.170&0.080&0.077&0.093&0.170\\ \hline
       9&  0.194&0.087&0.081&0.106&0.191\\ \hline
      10&  0.198&0.093&0.088&0.112&0.205\\ \hline\hline
        &0.322 &0.156& 0.182& 0.184& 0.223\\
        \hline
\end{tabular}
}
\caption{Experiment A}
\end{subtable}
\begin{subtable}{0.49\linewidth}\centering
\resizebox{0.8\linewidth}{!}{
\begin{tabular}{|c|c|c|c|c|c|}
\hline
$p$&$k=1$&$k=2$&$k=3$&$k=4$&$k=5$\\
\hline\hline
        1&0.015&0.015&0.014&0.014&0.014\\ \hline
   2&0.016&0.015&0.014&0.015&0.014\\ \hline
   3&0.016&0.015&0.015&0.015&0.020\\ \hline
   4&  0.017&0.015&0.015&0.016&0.021\\ \hline
   5&  0.021&0.016&0.015&0.017&0.021\\ \hline
   6&  0.021&0.017&0.018&0.018&0.023\\ \hline
   7&  0.021&0.018&0.018&0.019&0.024\\ \hline
   8&  0.023&0.018&0.018&0.019&0.025\\ \hline
   9&  0.025&0.019&0.019&0.019&0.026\\ \hline
   10&0.025& 0.019&0.019&0.019&0.026\\ \hline
   &  0.202&0.183&0.180&0.181&0.203\\ \hline
\end{tabular}
}
\caption{Experiment B}
\end{subtable}
\label{Table:experimentB}
\end{table}

\begin{figure}[!h]
    \begin{center}
    \begin{tabular}{cc}
        \includegraphics[width=1\textwidth]{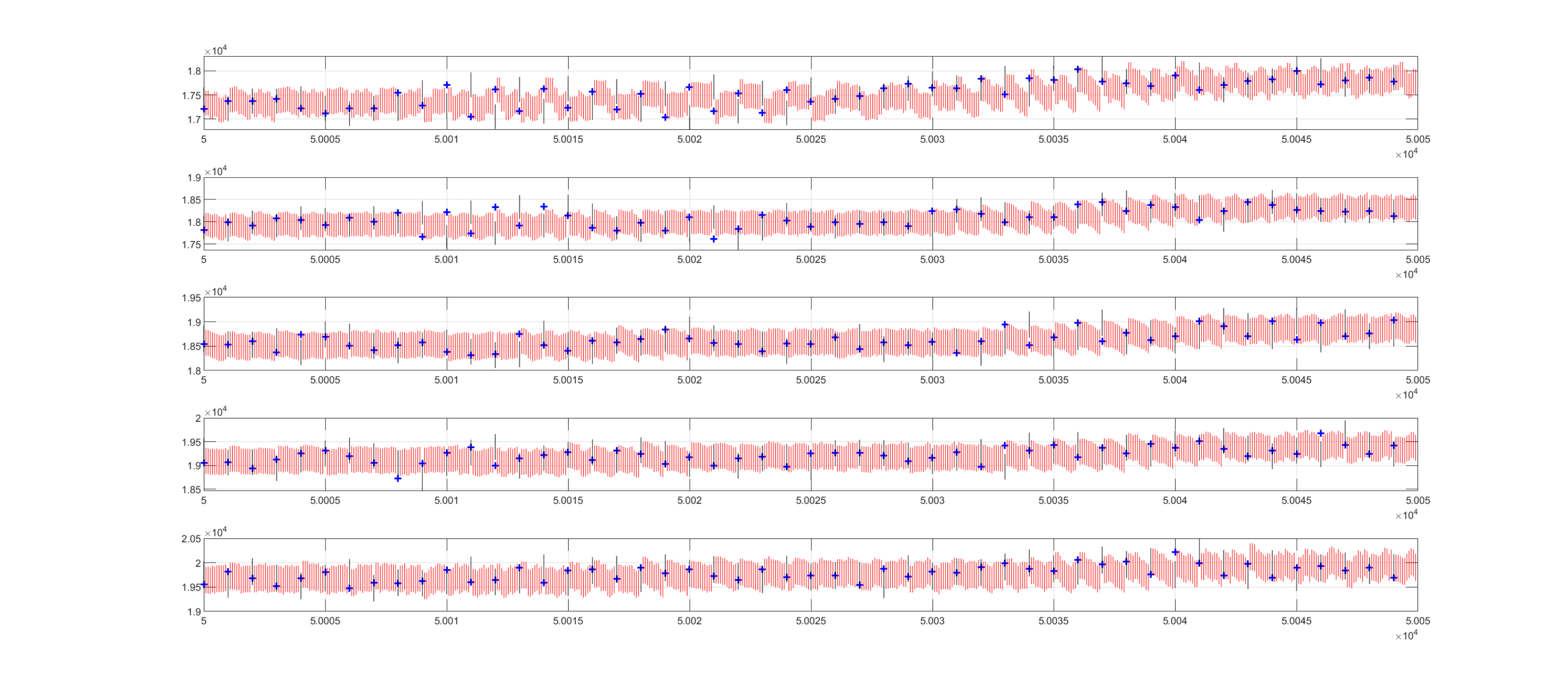}\\
       {Experiment A}\\
        \includegraphics[width=1\textwidth]{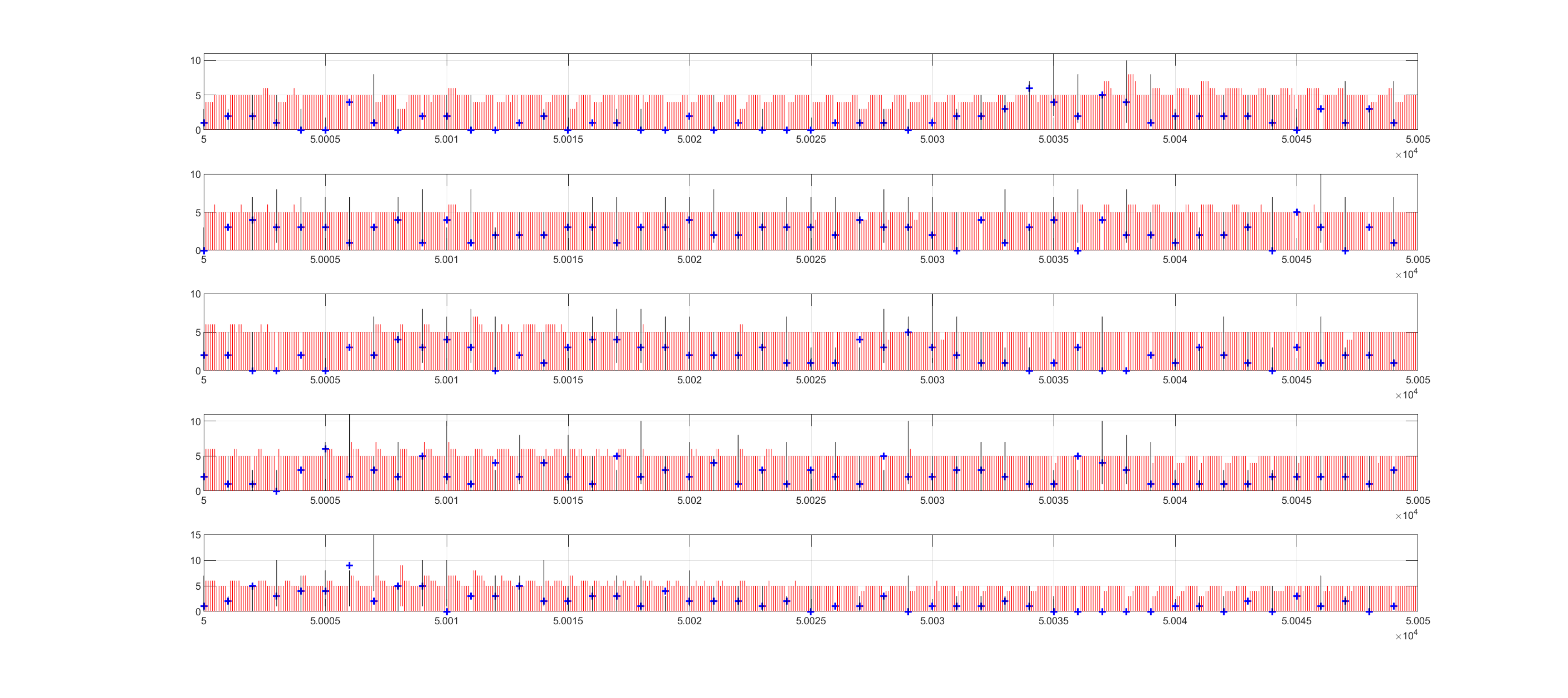}\\
        {Experiment B}
        \end{tabular}
        \end{center}
    \label{Fig:Pred_interval}
    \caption{Prediction on time interval $50,000\leq t\leq 50,050$. Top to bottom: locations. Blue crosses: actual states; 10 orange vertical segments:
in-between the crosses: 0.95-confidence intervals with prediction steps 1, 2, \ldots, 10; black vertical segments: confidence intervals yielded by trivial prediction. 
}
\label{Fig:Pred_interval}
\end{figure}



\clearpage
\section{Real Data illustrations}\label{sec:realData}

\subsection{Dataset}

\begin{figure}[!t]
\begin{minipage}{.25\textwidth}
        \centering
        \includegraphics[width=\linewidth]{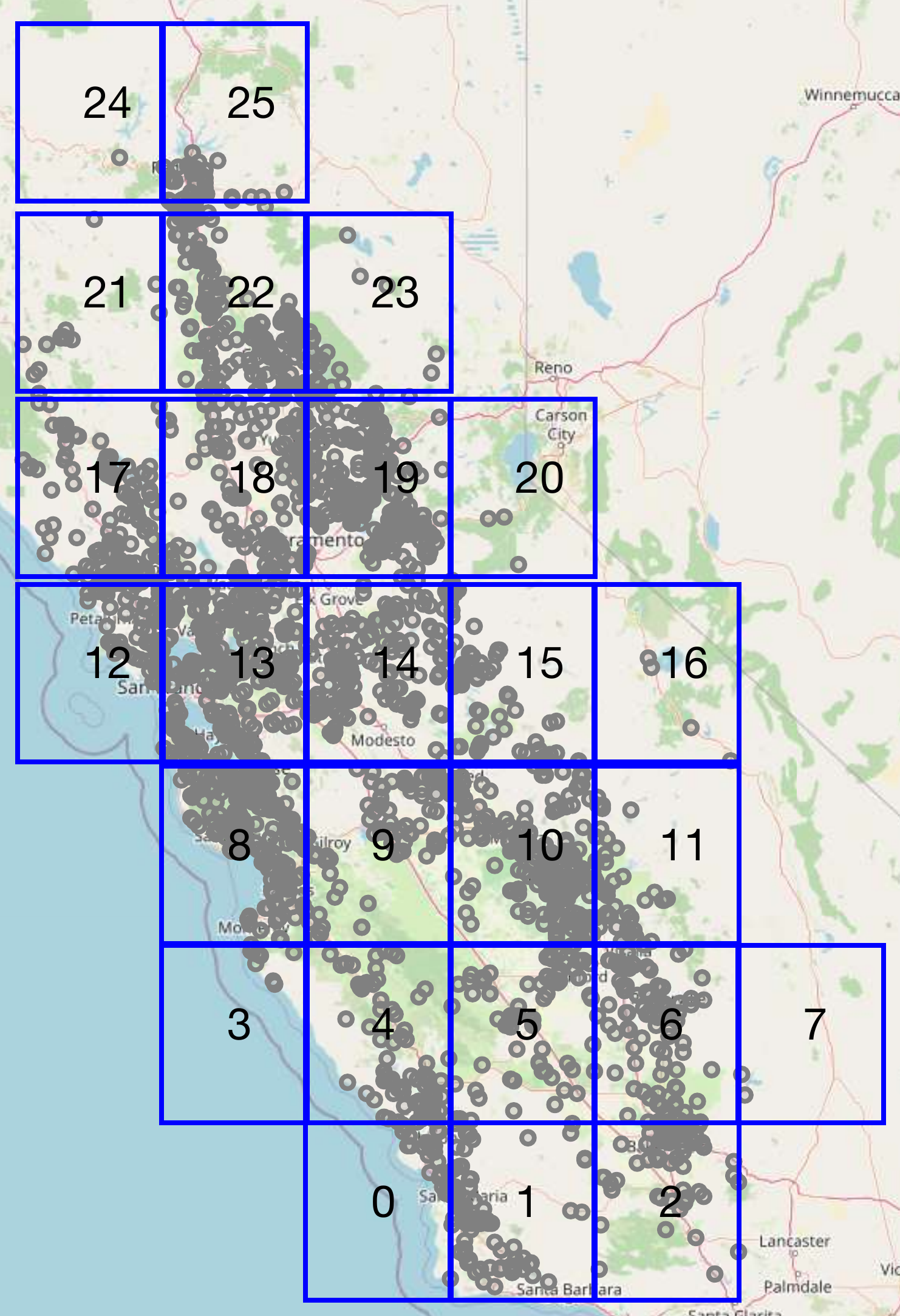}
        \captionsetup[subfigure]{justification=centering}
        \subcaption{Raw data and discretized space}
    \end{minipage}%
    \hspace{0.1in}
\begin{minipage}{.74\textwidth}
    \includegraphics[width=\linewidth]{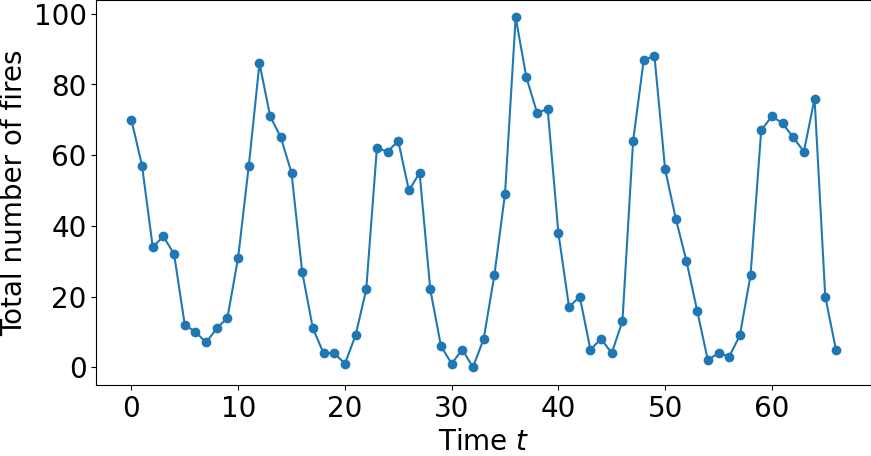}
    \subcaption{Total number of fires per month}
    
\end{minipage}
    \caption{{\small Figure (a) visualizes the wildfire data for training and test set, shown as scatters on the terrain map and spatial discretization for the Poisson process model. Figure (b) visualizes the total number of fire incidents over all locations at each unit time (month) $t$ (including both training and test time).}}
    \label{fig:wildfire_data}
\end{figure}

We apply the Poisson point process model to the wildfire dataset. The dataset contains wildfire occurrences in California from 06/2014 to 12/2019 at various locations. The dataset is retrieved from the California Public Utilities Commission\footnote{Website: \url{https://www.cpuc.ca.gov/wildfires/}}. We discretize the space containing the fire incidents into 26 disjoint rectangle cells (i.e., $L=26$); see Figure \ref{fig:wildfire_data} for the raw data (grey circles) and discretized cells (blue rectangles). Fire incidents are grouped by months, so each $[\zeta_t]_k$ denotes the number of wildfire incidents at location $k$ in month $t=1,\ldots,67$, after space discretization. Data from all locations and all except the last year (i.e., 55 months) are used for training, and data in the last year (i.e., 12 months) are for testing. We choose the memory depth $d=12$, assuming observations up to 12 months in the past may influence the present intensities.

\begin{figure}[!b]
\centering
\begin{minipage}{\textwidth}
    \includegraphics[width=\linewidth]{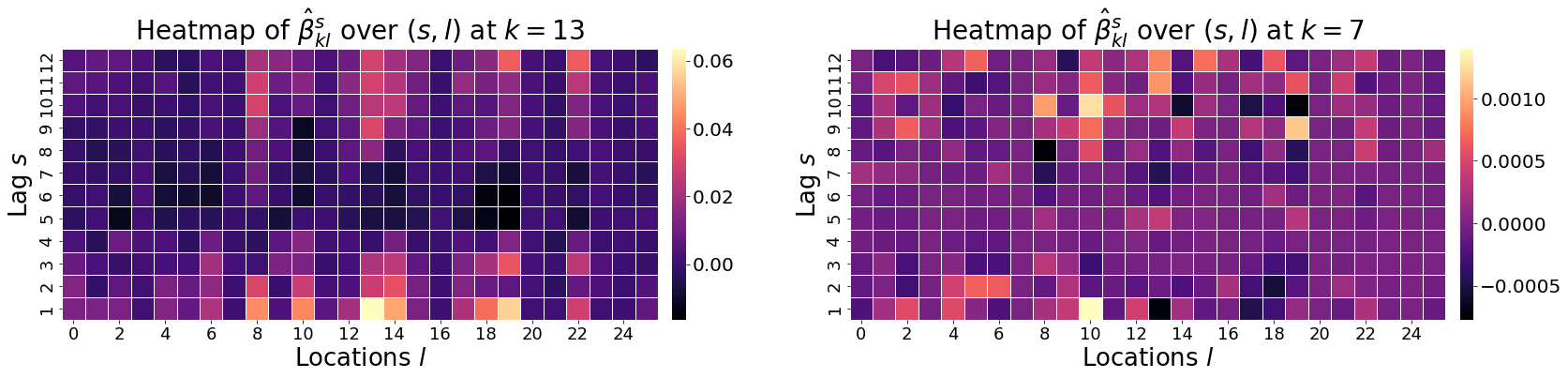}
    \vspace{-0.3in}
\end{minipage}
    \caption{{\small Visualization of estimated interaction parameters $\hat{\beta}^s_{kl}$ over neighbors $l=1,\ldots,L$ and lags $s=1,\ldots,d$. We select two locations $k$, where the left (resp. right) figure shows the heatmap of these interaction parameters at location 13 (resp. 7), with the highest (resp.) lowest fire incidents in training data. The estimated baseline intensity $\hat{\beta}^0_k$ at $k=13$ and $k=7$ are 0.0088 and 6.2$\times 10^{-5}$, respectively.}}
    \label{fig:param_est}
\end{figure}

\begin{figure}[!t]
    \begin{minipage}{0.305\textwidth}
        \includegraphics[width=\linewidth]{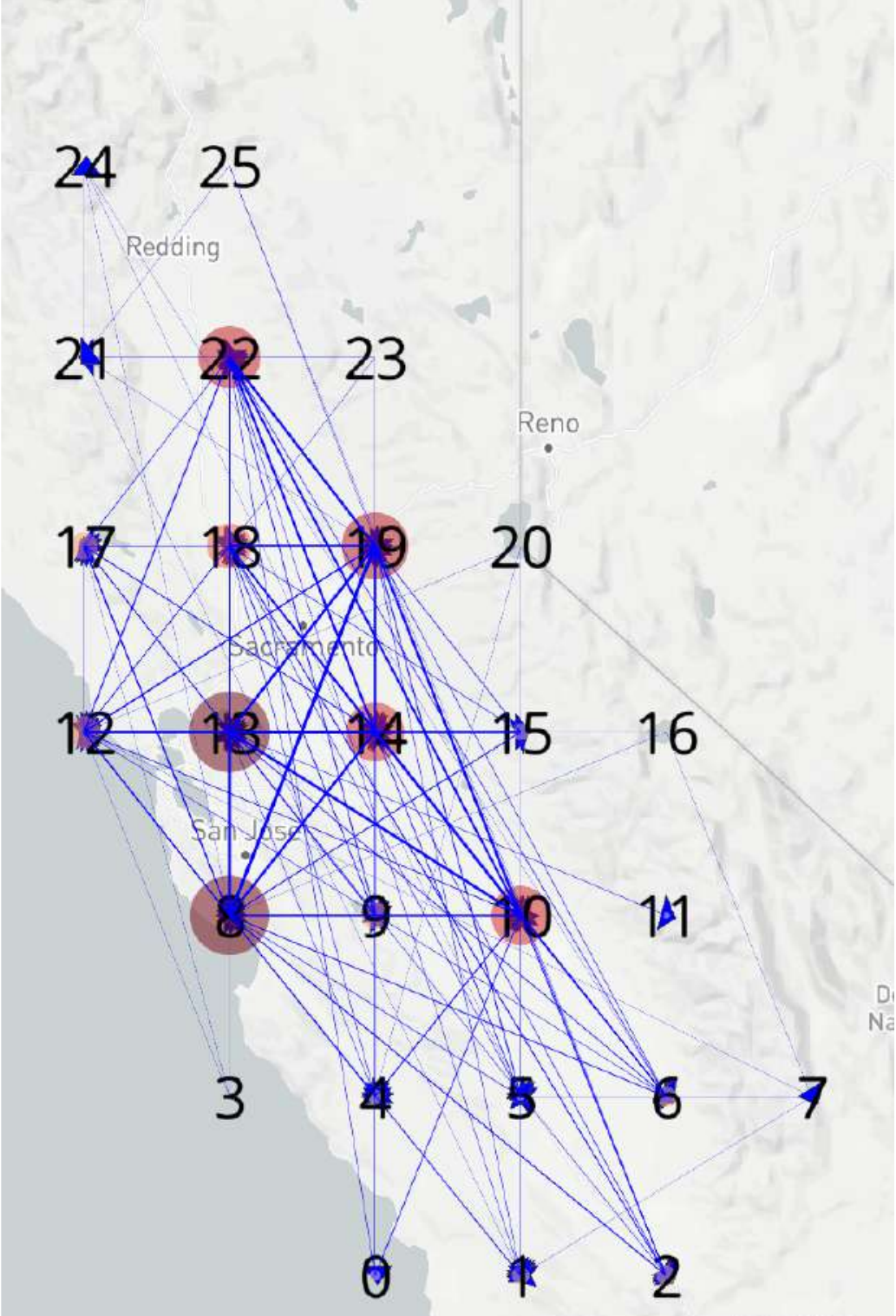}
        \subcaption{Lag $s=1$}
    \end{minipage}
    \begin{minipage}{0.3\textwidth}
        \includegraphics[width=\linewidth]{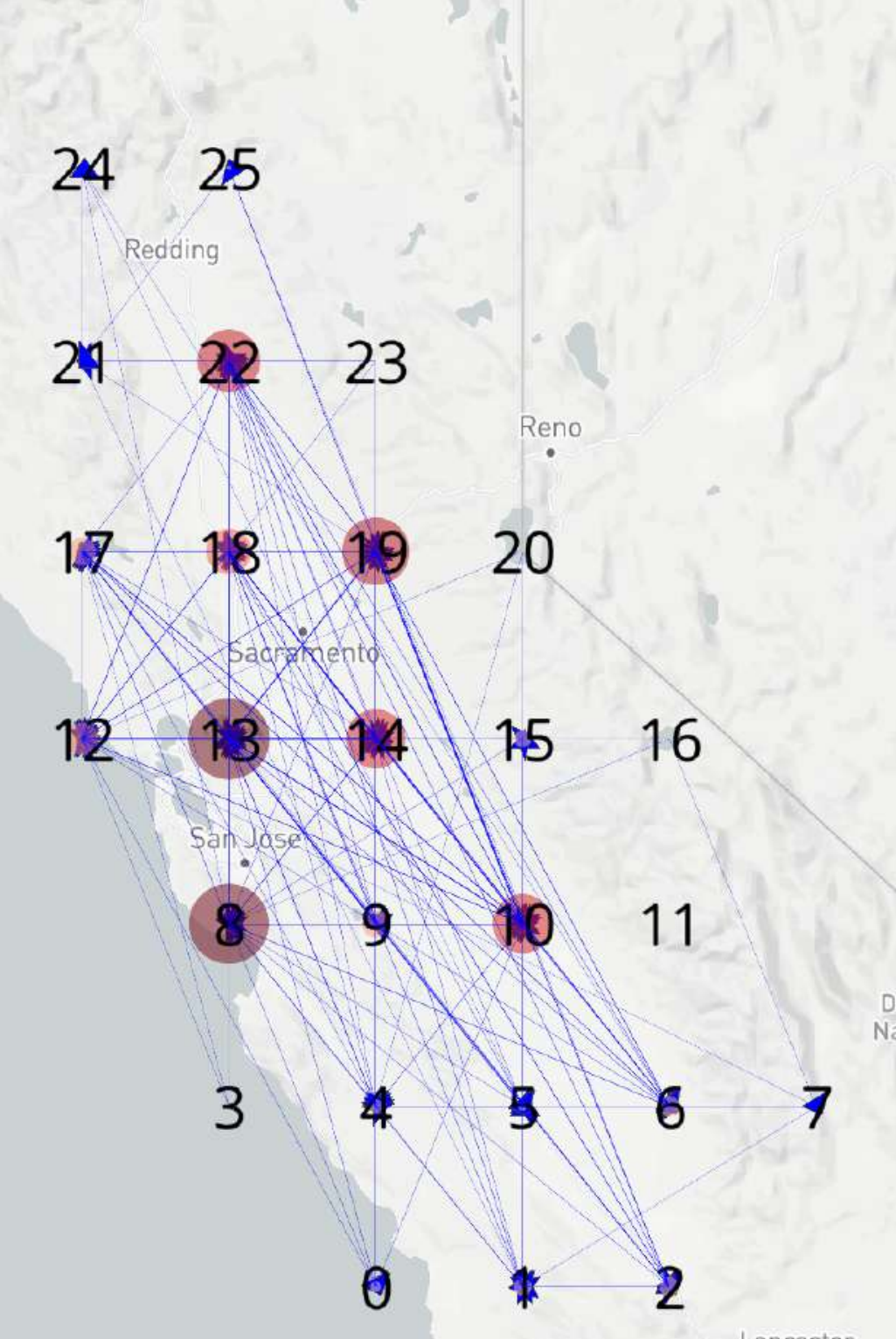}
        \subcaption{Lag $s=6$}
    \end{minipage}
    \begin{minipage}{0.38\textwidth}
        \includegraphics[width=\linewidth]{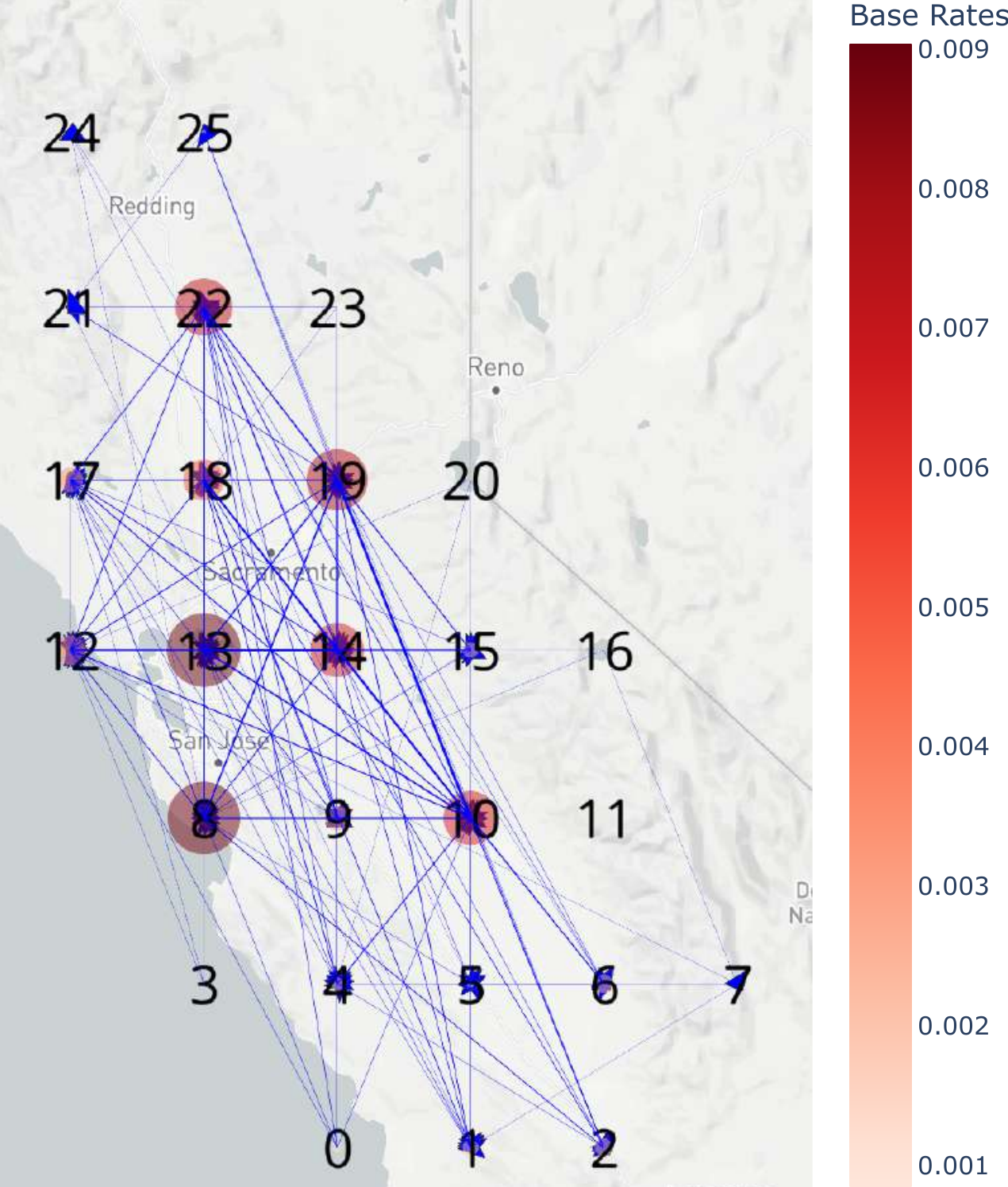}
        \subcaption{Lag $s=12$}
    \end{minipage}
    \caption{Overlaying base rates $\hat{\beta}^0_k$ and interaction parameters $\hat{\beta}^s_{kl}$ on a terrain map over different lag $s$. In each subfigure, the base rates are plotted as red solid circles, with color and size proportional to the value of $\hat{\beta}^0_k$. The interactions $\hat{\beta}^s_{kl}$ are plotted as blue-directed edges from location $l$ to location $k$, with width proportional to the magnitude.}
    \label{fig:param_est_on_map}
\end{figure}

\subsection{Model formulation and estimation}
Following the simulation examples, we assume the underlying point process is
\begin{equation}\label{fire_process}
[\zeta_t]_k\sim\mathrm{Poisson}\left(\beta^0_k+\sum_{s=1}^d\sum_{\ell=1}^L\beta^s_{k\ell}[\zeta_{t-s}]_\ell\right),\,t\geq1, 1\leq k\leq L.  
\end{equation}
The parameters $\beta=\{\beta^0_k, \beta^s_{kl}\}, \beta \in \mathbb R^{\kappa}$ are estimated as an optimal solution to the unconstrained least-square objective
\begin{equation}\label{fire_estimation}
    \min\limits_{x\in \mathbb R^{\kappa}}\left[{1\over 2N}\sum_{t=1}^N\sum_{k=1}^L\left([\zeta_t]_k-x^0_k-\sum_{s=1}^d\sum_{\ell=1}^Lx^s_{k\ell}[\zeta_{t-s}]_\ell\right)^2\right].
\end{equation}
During prediction, in order to ensure non-negativity of estimated rates in \eqref{fire_process}, we use $\max\{0,\beta^0_k+\sum_{s=1}^d\sum_{\ell=1}^L\beta^s_{k\ell}[\zeta_{t-s}]_\ell\}$ as the estimated rates. We note that unlike a constrained problem, which may require non-negativity in the entries of $\beta$, the unconstrained problem in \eqref{fire_estimation} allows for greater flexibility. Specifically, it can capture possibly negative spatial-temporal influences through the parameters $\beta^s_{kl}$.

\subsection{Evaluation metrics} 

We evaluate the model's performance on average and in specific instances. 

We consider several metrics for average performance. Below, $r_k$ denotes the observed average count (frequency) for a location $k$, and $\hat r_k$ denotes the conditional expectation for location $k$ based on the parameters estimated using training data from time 1 to $N$. We also consider a simulation-based test, $\hat{r}_k^{\text{simul}}$, which has been used in \cite[Table V]{juditsky2020convex} for Bernoulli processes; the rationale is we model the probability distribution of counts, and thus, can check whether or not our model can generate random instances with an empirical frequency that matches the observed frequency of the data. Specifically, we sample from a Poisson distribution with predicted intensity to generate replicas $[\hat{\zeta}_t]_{k}^{(i)}, i=1,\ldots, I$ for $I=100$. Finally, we also compare with a baseline using the seasonal prediction: in \eqref{seasonal_intensity}, the estimates $[\tilde{\zeta}_t]_k$ are seasonal averages (with a period of 12 months) of past samples $\{[\zeta_{t'}]_k: (t-t')\equiv 0 \ (\text{mod}~ 12), t' \leq N\}$. In our case, such predictions are meaningful due to seasonality in raw data, as shown in Figure \ref{fig:wildfire_data}(b). We thus compute the following quantities:
\begin{align}
     r_k &:=\frac{1}{T-N}\sum_{t=N+1}^T [\zeta_t]_k \label{true_freq} \\ 
    \hat{r}_k &:=\frac{1}{T-N}\sum_{t=N+1}^T \left(\hat{\beta}^0_k+\sum_{s=1}^d\sum_{\ell=1}^L\hat{\beta}^s_{k\ell}[\zeta_{t-s}]_\ell\right)\label{cond_intensity} \\
     \hat{r}_k^{\text{simul}} &:=\frac{1}{(T-N)I}\sum_{t=N+1}^T\sum_{i=1}^I [\hat{\zeta}_t]_{k}^{(i)} \label{cond_intensity_simul} \\
     \hat{r}_k^{\text{seasonal}} &:= \frac{1}{T-N}\sum_{t=N+1}^T [\tilde{\zeta}_t]_k \label{seasonal_intensity}.
\end{align}

\begin{table}[!t]
\centering
\caption{For wildfire data: the comparison of true average counts as in \eqref{true_freq} versus estimated average counts as in \eqref{cond_intensity}--\eqref{seasonal_intensity}. }
\label{poisson_table_fire}
\makebox[0pt][c]{\parbox{\textwidth}{%
\setstretch{1.35}
    \begin{minipage}[b]{0.48\hsize}\centering
        \resizebox{0.75\linewidth}{!}{\begin{tabular}{ | c |c|c|c|c|}
\hline
\textbf{Location} & \textbf{$r_k$}& \textbf{$\hat{r}_k$} & \textbf{$\hat{r}_k^{\text{simul}}$} & \textbf{$\hat{r}_k^{\text{seasonal}}$}\\ \hline
0 & 0.333 & 0.706 & 0.738 & 0.729 \\\hline
1 & 1.833 & 1.003 & 0.994 & 0.988 \\\hline
2 & 1.917 & 1.441 & 1.528 & 1.371 \\\hline
3 & 0.083 & 0.220 & 0.201 & 0.204 \\\hline
4 & 1.167 & 1.076 & 1.092 & 1.029 \\\hline
5 & 2.083 & 1.084 & 1.097 & 1.021 \\\hline
6 & 1.917 & 1.258 & 1.302 & 1.254 \\\hline
7 & 0.000 & 0.047 & 0.052 & 0.017 \\\hline
8 & 4.417 & 3.308 & 3.212 & 3.354 \\\hline
9 & 1.167 & 1.217 & 1.200 & 1.221 \\\hline
10 & 4.917 & 3.060 & 3.062 & 2.971 \\\hline
11 & 0.500 & 0.484 & 0.508 & 0.442 \\\hline
12 & 1.250 & 1.438 & 1.444 & 1.400 \\\hline
\end{tabular}}
         \subcaption{First 13 locations}
    \end{minipage}
    \hfill
        \begin{minipage}[b]{0.48\hsize}\centering
        \resizebox{0.75\linewidth}{!}{\begin{tabular}{ | c |c|c|c|c|}
\hline
\textbf{Location} & \textbf{$r_k$}& \textbf{$\hat{r}_k$} & \textbf{$\hat{r}_k^{\text{simul}}$} & \textbf{$\hat{r}_k^{\text{seasonal}}$}\\ \hline
13 & 5.333 & 4.000 & 4.123 & 3.996 \\\hline
14 & 2.167 & 2.334 & 2.357 & 2.392 \\\hline
15 & 0.917 & 0.763 & 0.757 & 0.746 \\\hline
16 & 0.083 & 0.063 & 0.065 & 0.038 \\\hline
17 & 1.167 & 1.567 & 1.658 & 1.529 \\\hline
18 & 2.583 & 2.527 & 2.514 & 2.388 \\\hline
19 & 2.667 & 3.832 & 3.899 & 3.596 \\\hline
20 & 0.000 & 0.055 & 0.057 & 0.050 \\\hline
21 & 0.167 & 0.274 & 0.278 & 0.208 \\\hline
22 & 2.250 & 2.691 & 2.734 & 2.683 \\\hline
23 & 0.333 & 0.196 & 0.177 & 0.204 \\\hline
24 & 0.083 & 0.049 & 0.054 & 0.021 \\\hline
25 & 0.333 & 0.367 & 0.386 & 0.350 \\\hline
\end{tabular}}
         \subcaption{Rest 13 locations}
    \end{minipage}
    \begin{center}
        \begin{minipage}[b]{0.4\hsize}\centering
        \resizebox{0.75\linewidth}{!}{\begin{tabular}{ | c | c |c|c| }
           \hline
            & \textbf{$\hat{r}_k$} & \textbf{$\hat{r}_k^{\text{simul}}$} & \textbf{$\hat{r}_k^{\text{seasonal}}$}\\
            \hline
           MAE & 0.421 &     0.424 &       0.425 \\\hline
        \end{tabular}}
        \subcaption{Mean Absolute Error (MAE) between $r_k$ and the estimates}
    \end{minipage}
    \end{center}
}}

\end{table}

\subsection{Results}

We first visualize estimated parameters, both as heatmaps over lag $s=1,\ldots,d,$ in Figure \ref{fig:param_est} and edges overlaid on a terrain map in Figure \ref{fig:param_est_on_map}. Figure \ref{fig:param_est} shows that at location $k=13$ with the most fire incidents, the estimated $\beta^s_{kl}$ are much higher in magnitude than those at location $k=7$ (with the least fire incidents). The large magnitude parameters indicate significant spatial-temporal influence received by location $k$ from its neighboring and past fire incidents. Such strong influences are also shown in Figure \ref{fig:param_est_on_map}, visualized as directed edges. Meanwhile, from both figures, we see that values of $\hat{\beta}^s_{kl}$ tend to be smaller as the lag $s$ increases (especially at $s=1$ versus $s=6$), indicating a temporally decaying influence from historical fire incidents.

\begin{figure}[!b]
    \centering
    \includegraphics[width=\linewidth]{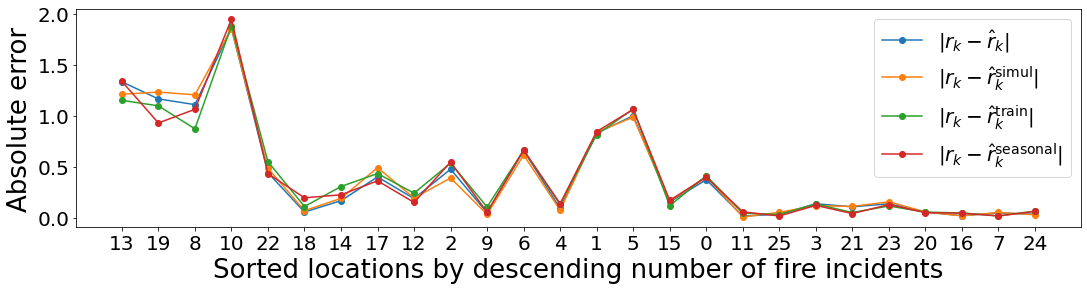}
    \caption{Absolute error of metrics shown in Table \ref{poisson_table_fire}. The locations on the x-axis are sorted by the observed number of fire incidents in the training data.}
    \label{sorted_AE}
\end{figure}

\begin{table}[!t]
    \centering
    \caption{For wildfire data, empirical 0.95-confidence interval for $p$-step-ahead prediction. We report top-2, middle-2, and bottom-2 locations with the fire incidents level that are high, medium, and low in training data. The second-to-bottom row labeled with {\it persistent prediction} refers to the prediction using the previous observation$[\zeta_{t-1}]_k$, assuming tomorrow will be the same as today. The bottom row predicts using seasonal average $[\tilde{\zeta}_t]_k$ defined in \eqref{seasonal_intensity}.}
    \label{tab:non_coverage}
    \centering
    \resizebox{0.75\linewidth}{!}{
    \begin{tabular}{|c|c|c|c|c|c|c|}
        \hline
    $p$&$k=13$&$k=19$&$k=4$&$k=1$&$k=7$&$k=24$\\
    \hline\hline
1                    &  0.167 &  0.000 &  0.000 &  0.167 &  0.0 &  0.000 \\\hline
2                    &  0.240 &  0.000 &  0.000 &  0.182 &  0.0 &  0.000 \\\hline
3                    &  0.246 &  0.002 &  0.001 &  0.200 &  0.0 &  0.000 \\\hline
4                    &  0.242 &  0.002 &  0.002 &  0.222 &  0.0 &  0.000 \\\hline
5                    &  0.267 &  0.000 &  0.000 &  0.250 &  0.0 &  0.000 \\\hline
6                    &  0.301 &  0.000 &  0.007 &  0.285 &  0.0 &  0.000 \\\hline
7                    &  0.333 &  0.000 &  0.004 &  0.333 &  0.0 &  0.000 \\\hline
8                    &  0.400 &  0.000 &  0.012 &  0.400 &  0.0 &  0.000 \\\hline
9                    &  0.500 &  0.000 &  0.014 &  0.250 &  0.0 &  0.000 \\\hline
10                   &  0.333 &  0.000 &  0.018 &  0.333 &  0.0 &  0.000 \\\hline
11                   &  0.000 &  0.000 &  0.029 &  0.000 &  0.0 &  0.000 \\\hline
12                   &  0.000 &  0.000 &  0.103 &  0.000 &  0.0 &  0.000 \\\hline
Persistent pred, $p=1$ &  0.417 &  0.167 &  0.167 &  0.250 &  0.0 &  0.083 \\\hline
Seasonal average pred, $p=1$ &  0.083 &  0.083 &  0.000 &  0.167 &  0.0 &  0.083 \\\hline
    \end{tabular}
    }
\end{table}

\begin{figure}[!b]
        \centering
        \includegraphics[width=\linewidth]{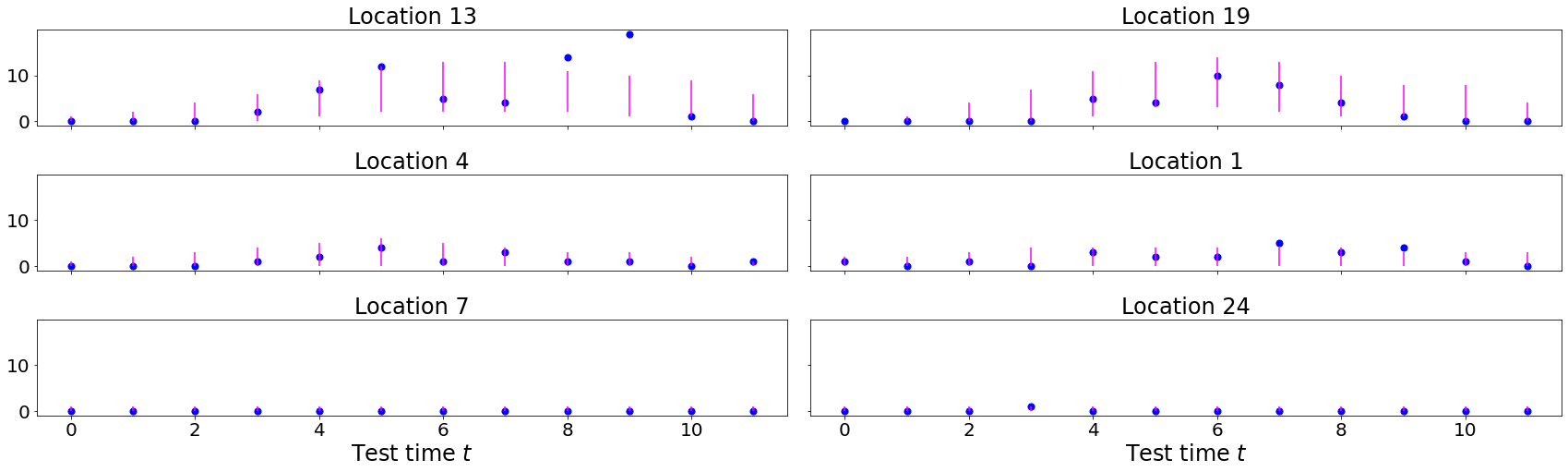}
    \caption{Instance-based ``online'' prediction intervals on test data at locations selected in Table \ref{tab:non_coverage}: observations are blue dots, and intervals are in magenta; we note that the prediction interval achieves good coverage probability with a reasonable width.}
     \label{fire_interval}
\end{figure}

We then compare the performance of our model against baselines defined in \eqref{seasonal_intensity}. Table \ref{poisson_table_fire}(c) shows that, on average, the Mean Absolute Error (MAE) of estimated rates by our model is lower than that of the baselines. Upon sorting all locations by descending number of fire incidents, we see in Figure \ref{sorted_AE} that our method yields very similar performance in most locations.

We lastly compare on an instance level the non-coverage of prediction intervals as considered in simulation (i.e., Table \ref{Table:experimentB}). For instance-based performance, we report the $p$-step ahead ($p = 1, \ldots, 12$)  interval non-coverage (similar to Table \ref{Table:experimentB} and Figure \ref{Fig:Pred_interval} for simulation examples). Here, Table \ref{tab:non_coverage} shows $p$-step ahead coverage at different prediction horizons for the wildfire data, where the non-coverage tends to increase as the model predicts further into the future. We observe that our model incurs a lower non-coverage than the persistent prediction and the seasonal average prediction at $p=1$ (i.e., one-step ahead prediction). Specifically, our model incurs a non-coverage exceeding the target 0.05 only at 2 out of 6 locations, whereas the other baselines incur non-coverage at 5 or 4 out of 6 locations. Figure \ref{fire_interval} visualizes prediction intervals when $p=1$ at corresponding locations, where intervals are also adaptive in widths at different locations and test times with a reasonable width.

\section{Conclusion}

We have introduced a computational method to estimate and quantify the uncertainty in the Generalized Generalized Linear Model (GGLM) for modeling spatio-temporal data. Our method provides a parameter recovery guarantee as well as uncertainty quantification through a set of concentration inequalities. To assess the performance of our method, we conducted numerical simulations and applied the method to a real-world problem of wildfire prediction.

\bibliography{ReferencesChP}

\bibliographystyle{plain}

\end{document}